\documentclass[
aps,
prb,
reprint,
superscriptaddress,
longbibliography,
nofootinbib
]{revtex4-2}

\usepackage{amsmath,amsfonts,graphicx,color,bbm,tikz,bm,setspace,overpic}
\usepackage{graphicx}
\usepackage{hyperref}
\usepackage{float}
\usepackage{bbold}
\usepackage{subfig}
\usepackage{multirow}

\begin{document}

\title{Hidden Ising models from the generalized Yang-Baxter equation}

\author{Akash Sinha}
\email{akash26121999@gmail.com}
\affiliation{Department of Physics, School of Basic Sciences,\\ Indian Institute of Technology, Bhubaneswar, 752050, India}

\author{Somnath Maity}
\email{somnathmaity126@gmail.com}
\affiliation{Department of Physics, School of Basic Sciences,\\ Indian Institute of Technology, Bhubaneswar, 752050, India}

\author{Pramod Padmanabhan}
\email{pramod23phys@gmail.com}
\affiliation{Department of Physics, School of Basic Sciences,\\ Indian Institute of Technology, Bhubaneswar, 752050, India}

\author{Vladimir Korepin}
\email{vladimir.korepin@stonybrook.edu}
\affiliation{C. N. Yang Institute for Theoretical Physics, \\ Stony Brook University, New York 11794, USA}

\begin{abstract}
We introduce a one dimensional spin $\frac{1}{2}$ Hamiltonian with multi-site interactions, but still local. The algebra of its Hamiltonian densities resembles that of the transverse field Ising model. Using this fact we show that its spectrum is free-fermionic but with a huge degeneracy for each level. The source of the degeneracy is a set of local conserved quantities that act like a classical background field for the quantum system. The thermodynamics of this system is contrasted with the standard Ising model. At the gapless points in the energy spectrum, we show that this system can be derived from the quantum inverse scattering method adapted to a multi-site generalization of the Yang-Baxter equation as introduced by E. Rowell and Z. Wang. The $R$-matrix is constructed using generators of extraspecial 2-groups. This helps us extract all the conserved charges and lay the framework for a general mechanism to generate such multi-site interaction spin systems that are transverse field Ising models under the hood. A remark on how to obtain P. Fendley's free-fermion in disguise models in this formalism is also included. 
\end{abstract}

\maketitle

\tableofcontents 

\section{Introduction}
\label{sec:Introduction}
Exactly solvable models are essential to understand several quantum many-body phenomena. Among them quantum spin chains are the most widely studied systems. They are usually described by Hamiltonian densities that are either local or quasi-local, i.e. as operators, they have support on nearest-neighbors or a few more consecutive sites on the one dimensional chain. The algebra of these Hamiltonian densities can provide insights into the physical behavior of the spin chains. In particular, the spectrum of these systems can be obtained in a representation independent way using these local algebras. Thus a large class of spin chains will have similar spectrum provided their Hamiltonian densities obey the same local algebra. In the last few years this approach has been used to identify local algebras that lead to {\it free-fermionic} systems \cite{Onsager-1944,Kaufman-1949,LSM-1964}. The latter are Hamiltonians that can be written as bilinears in Dirac or Majorana fermions. Historically, many standard spin chains, such as the {\it transverse field Ising model} \cite{PFEUTY197079}, the anisotropic $XY$-model and the simpler $XX$-model \cite{LIEB1961407} and the relatively more recent {\it cluster Ising models} \cite{Hans-2001,son2011quantum} are known to be free-fermionic using a standard {\it Jordan-Wigner transformation} \cite{JW-original}. Only recently, more exotic and non-trivial examples, known as {\it free-fermions in disguise}, were initiated by P. Fendley \cite{fendley2019free}. These are spin systems, with multi-site interactions, that do not become free-fermionic with a naive application of the Jordan-Wigner transformation. Their Hamiltonian densities satisfy a particular local algebra, \eqref{eq:FFD-algebra} . Since its inception, a few generalizations in different directions have been proposed. These include multi-spin and parafermionic generalizations, \cite{Alcaraz-2020-1,Alcaraz-2020-2,Poszgay-2021}, embedding these systems in an algebraic context \cite{Miao-2022}, and in open systems \cite{fukai2026dissipativefreefermionsdisguise}.
In a another relevant direction, the general problem of devising prescriptions to detect free-fermion Hamiltonians among spin Hamiltonians has been formulated using ideas from graph theory
\cite{elman2021free, Chapman_2020} and simplicial homology \cite{Ogura_2020}. The resulting models encompass several different local algebras of Hamiltonian densities that go beyond the algebra obeyed by free-fermions in disguise. Discrete-time versions of free-fermions in disguise models have also been constructed \cite{vona2025exactrealtimedynamics,Sz_sz_Schagrin_2026}, generalizing match-gate circuits used in the simulation of free-fermion systems \cite{Terhal_2002,Jozsa_2008}. 

In this work, we take a step back and analyze the local algebra underlying the transverse field Ising model (TFIM). The Hamiltonian of interest is given by
\begin{eqnarray}\label{eq:Ham_TFIM}
    H_{\rm TFIM}=-g\sum_jZ_j-\sum_j X_jX_{j+1},
\end{eqnarray}
where $X_j,Z_j$ are the spin-1/2 Pauli matrices, acting non-trivially only on site $j$. The local Hamiltonian densities satisfy the relations
\begin{eqnarray}\label{eq:isingzxrelations}
    Z_j\left(X_kX_{k+1}\right)=\begin{cases}
        -\left(X_kX_{k+1}\right)Z_j,\quad j=k,k+1,\nonumber\\
        \left(X_kX_{k+1}\right)Z_j,\qquad j\neq k,k+1.
    \end{cases}
\end{eqnarray}
At $g=1$, the system undergoes a quantum phase transition and becomes critical. This critical point belongs to the $(1+1)$-D Ising universality class and is described by a \textit{conformal field theory} (CFT) with central charge $c=1/2$. When written in terms of the Majorana modes
\begin{eqnarray}\label{eq:JW1-transformation}
    &\gamma_{2j-1}=\left(\prod\limits_{k=1}^jZ_j\right)X_j,\quad \gamma_{2j}=\left(\prod\limits_{k=1}^jZ_j\right)Y_j,\nonumber\\
    &\{\gamma_j,\gamma_k\}=2\delta_{j,k},
\end{eqnarray}
the Hamiltonian in \eqref{eq:Ham_TFIM} assumes the free-fermionic form 
\begin{eqnarray}\label{eq:Ising_Majorana}
    H_{\rm TFIM}={\rm i}g\sum_j\gamma_{2j-1}\gamma_{2j}+{\rm i}\sum_{j}\gamma_{2j}\gamma_{2j+1}.
\end{eqnarray}
The above mapping between the spins and the fermions is known as the Jordan-Wigner (JW) transformation\footnote{While the transformation in \eqref{eq:JW1-transformation} holds for one dimension, there exists several higher dimensional generalizations \cite{Fradkin-JW-1989,Wang-JW-1991, Zanelli-JW-1993, Ortiz-JW-2001, Oritz-JW-2012, Backens-JW-2019,Verstraete_2005,CHEN2018234,Tantivasadakarn_2020}. Even in one dimensions the JW transform is not unique and many variants can be generated in an algorithmic manner \cite{Bonsai-JW-2023}. }. For future convenience, we denote the above JW transformations by the shorthand notation $\Phi_{{\rm JW}_1}$. Traditionally, the Hamiltonian \eqref{eq:Ising_Majorana} is diagonalized by performing a Fourier transformation to momentum space, followed by a Bogoliubov transformation that diagonalizes the resulting quadratic form. This a generic feature of the quadratic fermion Hamiltonians: they can, in principle, be diagonalized by a suitable canonical transformation, reducing it to a set of independent modes. The spectrum and dynamics are thereby exactly solvable. Though explicit expressions for the energies may still be nontrivial in practice, they usually take the form
\begin{eqnarray}
    E = \pm\epsilon_1 \pm \epsilon_2 \cdots \pm\epsilon_N,
\end{eqnarray}
on a $N$-site chain.

The local algebra underlying the TFIM Hamiltonian, \eqref{eq:Ham_TFIM} is given by
\begin{eqnarray}\label{eq:Ising_algebra}
&\left[h^z_j,h^z_k\right]=0=\left[h^{xx}_j,h^{xx}_k\right],~\left[h^z_j,h^{xx}_k\right]=0,~j\neq k,k+1, \nonumber\\&\{h^z_{j+1},h^{xx}_{k}\}=0=\{h^z_j,h^{xx}_{j}\},\nonumber\\&\left(h^z_j\right)^2=1=\left(h^{xx}_j\right)^2.
\end{eqnarray}
The operators in \eqref{eq:isingzxrelations} with $h^z_j=Z_j$ and $h^{xx}_j=X_jX_{j+1}$, satisfy this algebra. We will call this the {\it Ising exchange algebra} for the Hamiltonian densities of the TFIM. As shown in \cite{minami2016solvable}, the above algebraic structure alone guarantees a free-fermionic spectrum, irrespective of its actual realization in terms of spin-1/2 operators. 
To be precise, one can construct a set of fermionic operators $\{\varphi_j\}$, with $j=1,\cdots,N$. They satisfy the \textit{canonical anti-commutation relation} (CAR)
\begin{eqnarray}\label{eq:Minami's-Majorana}
    &  \varphi_{2j-1}=e^{{\rm i}(j-1)\pi}\left(\prod\limits_{k=1}^jh_k^z h_k^{xx}\right)h_j^z,\nonumber\\
    &\varphi_{2j}=e^{{\rm i}\left(j-\frac{1}{2}\right)\pi}\left(\prod\limits_{k=1}^jh_k^z h_k^{xx}\right)h_j^zh_j^{xx}, & \nonumber \\
     & \{\varphi_j,\varphi_k\}=2\delta_{j,k}. &
\end{eqnarray}
In terms of these variables the Hamiltonian becomes quadratic. Consequently, the Hamiltonian can be trivially diagonalized and the system becomes exactly solvable. Notably, such theories are distinct from Fendley's free fermions in disguise (FFD)\cite{fendley2019free} models, which do not admit such decomposition \cite{elman2021free}.

The solvability of a free-fermionic theory is generally lost once interactions are introduced in the system. The associated Hamiltonian is no longer quadratic in the fermionic operators, and as a result, the conventional momentum-mode diagonalization procedure typically breaks down. Nonetheless, there are certain systems, which possess hidden free-fermionic structures, even though the local Hamiltonian densities may initially appear to be interacting. In this work we construct precisely such systems as representations of the Ising exchange algebra, \eqref{eq:Ising_algebra}. These realizations have support on multiple sites of the one dimensional lattice, in both the spin basis and in the Majorana fermion basis\footnote{The number of spins is half the number of Majorana fermions from the JW transform in \eqref{eq:JW1-transformation}.}. Thus a superficial glance at the Hamiltonian will suggest that the system is interacting in both these realizations. Nevertheless, we will show that this is in fact not the case and the system is effectively a TFIM on a much reduced lattice size. This conclusion is drawn by following a 3-step procedure:
\begin{enumerate}
    \item We begin with a Hamiltonian with multi-site interaction in the spin basis ($X$, $Y$, $Z$). The particular Hamiltonian we write down acts on a one dimensional chain with $3N$ sites. The local Hamiltonian densities, satisfying the Ising exchange algebra \eqref{eq:Ising_algebra}, have non-trivial support on 5 sites. This Hamiltonian is then mapped to a Majorana fermion system with $6N$ Majoranas ($\gamma$), using the standard JW transform as in \eqref{eq:JW1-transformation}. The Hamiltonian densities are now quartic in these $\gamma$'s.
    \item Next we identify that the local Hamiltonian densities of this Majorana fermion Hamiltonian can be rewritten as a bilinear using two flavors of Majoranas, $a_j$ and $b_j$. The index $j$ now runs over $2N$ Majoranas. The new CAR algebra satisfied by the $a_j$'s and $b_j$'s has a center\footnote{The center of an algebra is the set of elements that commutes with every other element of the algebra.} generated by local operators $\mathcal{C}_j$'s, which are mutually commuting. The dimension of this center is extensive in the lattice size. The Majorana Hamiltonian in $a_j$'s and $b_j$'s block diagonalizes in the simultaneous eigenspace of the $\mathcal{C}_j$'s. In each of these blocks the Hamiltonian resembles a TFIM written in terms of one of the flavors of the effective Majoranas $a_j$ or $b_j$.
    \item Finally, we perform a standard JW transform on each of these blocks to obtain the effective TFIM with the spin variables $\tilde{X}$, $\tilde{Y}$ and $\tilde{Z}$. Each eigenvalue comes with a huge degeneracy that is now accounted for by the local conserved quantities $\mathcal{C}_j$'s.
\end{enumerate}
It is important to note that the Ising exchange algebra in \eqref{eq:Ising_algebra} does not have a center in general. The center generated in our construction can be attributed to the fact that we have identified particular representations of this algebra that allows this structure. We further note that these central elements cannot be generated by the $\varphi$ variables in \eqref{eq:Minami's-Majorana}.

We then show that this Hamiltonian is not an isolated occurrence. Instead, such multi-site models can be algorithmically constructed using the \textit{quantum inverse-scattering method} (QISM) \cite{Korepin1993QuantumIS} adapted to a multi-site generalization of the Yang-Baxter equation, known as the \textit{generalized Yang–Baxter equation} (gYBE) \cite{rowell2010extraspecialtwogroupsgeneralizedyangbaxter,Padmanabhan2019QuantumES}. The $R$-matrices used for this purpose have multiple indices and are constructed using generators of {\it extraspecial 2-groups}. This proves the integrability of the multi-site Ising model extending our earlier construction of the TFIM from the quantum inverse scattering formalism \cite{Sinha:2025wqf}. This also helps us systematically write down all the conserved quantities of this model using the {\it boost operator formalism} \cite{loebbert2016lectures}. 

The contents are laid out as follows. The paper is split into two parts. The first part illustrates the above mechanism with a specific multi-site Hamiltonian. The model, in the spin basis, is defined and certain symmetries are described in Section \ref{sec:The multi-site Hamiltonian}. The spectrum, with different boundary conditions, is discussed in detail in Section \ref{sec:spectrum}. Then the partition function and the associated thermodynamics are discussed in Section \ref{sec:thermodynamics}.  Following this, Section \ref{sec:integrability} discusses the integrability of the above Hamiltonians. The second part of the paper explores the construction of a broader class of such multi-site Hamiltonians. These constitute the Sections \ref{sec:Long-range Hamiltonians} and \ref{sec:gYBE-HFF}. In Section \ref{sec:Long-range Hamiltonians} we use our existing framework to generate Hamiltonians which have longer range of interactions. Subsequently, in Section \ref{sec:gYBE-HFF}, we develop a more general framework for constructing such Hamiltonians and present several explicit examples. A short conclusion summarizing the main results and scope for future work is presented in Section \ref{sec:outlook}. Several appendices elaborate on many missing details from the main text.

\section{The multi-site local Hamiltonian}
\label{sec:The multi-site Hamiltonian}
\begin{figure}[H]
    \centering
    \includegraphics[width=0.95\linewidth]{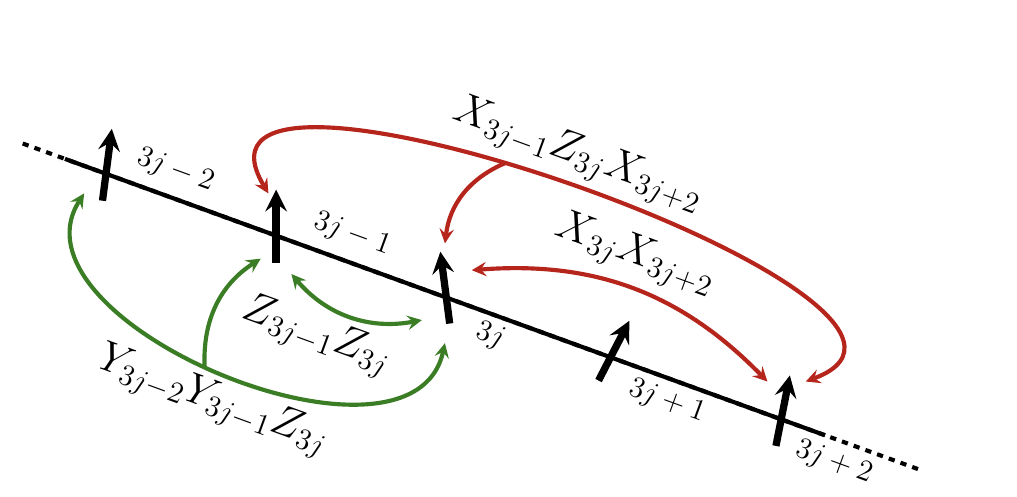}
    \caption{Action of the Hamiltonian density of \eqref{eq:hzx}.}
    \label{fig:6-3-H-action}
\end{figure}

Consider a spin-1/2 system defined over an one-dimensional lattice of $3N$-sites ($N\in\mathbb{Z}^+$), with the Hamiltonian given by
\begin{eqnarray}\label{eq:(6,3)-H}
    H=g\sum_{j=1}^{N}h^z_j+\sum_{j=1}^{N}h^{xx}_j,
\end{eqnarray}
where the local Hamiltonian densities are given by
\begin{eqnarray}\label{eq:hzx}
    &&h^z_j=\zeta Z_{3j-1}Z_{3j}+\eta Y_{3j-2}Y_{3j-1}Z_{3j},\nonumber\\
    &&h^{xx}_j=\tilde{\zeta}X_{3j}X_{3j+2}+\tilde{\eta}X_{3j-1}Z_{3j}X_{3j+2},
\end{eqnarray}
with $\zeta$, $\eta$, $\tilde{\zeta}$ and $\tilde{\eta}$ being real parameters. As in the standard TFIM \eqref{eq:Ham_TFIM}, the $h_j^z$ terms act on disjoint regions, while neighboring $h_j^{xx}$ terms have overlapping supports. For this reason the parameter $g$ is analogous the magnetic field strength and is introduced to control the phase structure of the model, analogous to what is seen in the TFIM case, \eqref{eq:Ham_TFIM}. We impose periodic boundary conditions on the local operators ${\cal O}_{3N+j}\simeq {\cal O}_j$, where ${\cal O}=X,Y,Z$. Note that, the above Hamiltonian densities obey the Ising exchange algebra \eqref{eq:Ising_algebra}, when
\begin{eqnarray}
    \zeta^2+\eta^2=\tilde{\zeta}^2+\tilde{\eta}^2=1.
\end{eqnarray}
The action of the Hamiltonian densities, \eqref{eq:hzx} is shown in Figure \ref{fig:6-3-H-action}.
A naive application of the JW transformation \eqref{eq:JW1-transformation} turns this into a seemingly interacting fermion system. However, as we shall explain below, this system is free-fermionic, and hence effectively a TFIM, after multiple transformations\footnote{Multi-site Ising models, especially with 4-site interactions, have been discussed in the past \cite{Wu-multi-ising-1971,wegner2014dualitygeneralizedisingmodels, Wegner-multi-Ising-1971} in the context of generalized Kramers-Wannier duality. A more recent work with a generalized Ising exchange algebra can be found in \cite{Alcaraz_2023}. As the Hamiltonian densities in these models obey a different algebra, we do not expect them to be related to the models presented in this work.}.
An immediate consequence of the latter is that this Hamiltonian becomes a TFIM on a lattice smaller in size than the one we started with. This implies that the spectrum of this Hamiltonian is exponentially degenerate. The presence of such a large degeneracy can be understood in terms of the conserved elements
\begin{eqnarray}\label{eq:spin-central}
            &&{\cal I}_{j}=\zeta Z_{3j-2}+\eta X_{3j-2}X_{3j-1},\nonumber\\
            &&\tilde{\cal I}_{j}=\tilde{\eta} Z_{3j}+\tilde{\zeta} X_{3j-1}X_{3j}.
        \end{eqnarray}
Note that, the above local conserved quantities closely resemble the local Ising densities on sublattices. All of them commute with the Hamiltonian $H$ and also commute with each other
\begin{eqnarray}
&\left[{\cal I}_j,{\cal I}_k\right]=0=\left[\tilde{\cal I}_j,\tilde{\cal I}_k\right],\qquad \left[{\cal I}_j,\tilde{\cal I}_k\right]=0,\nonumber\\
&\left[{\cal I}_j,H\right]=0=\left[\tilde{\cal I}_j,H\right],
\end{eqnarray}
with $j,k=1,\cdots,N$. As a result, the Hilbert total space splits into $2^{2N}$ sectors, labeled by the eigenvalues of $\{{\cal I}_j,\tilde{\cal I}_j\}$. We shall show that, despite the exponentially large number of sectors, only two of them are spectrally inequivalent. The rest of the sectors are isospectral with either one of them. This leads to the exponentially large degeneracy of order $2^{2N-1}$ in the global spectrum. 

It should also be noted that the operator $Y_{3j-2}$, for each $j$, trivially commutes with the Hamiltonian. However, these operators do not commute with the $\mathcal{I}_j$ central elements in \eqref{eq:spin-central} but does commute with the $\tilde{\mathcal{I}}_j$ elements and hence we can also block diagonalize the Hamiltonian using these two sets of mutually commuting conserved operators. Furthermore, the appearance of $Y_{3j-2}$ ensures that the Hamiltonian $H$ in \eqref{eq:(6,3)-H} enjoys the $\mathbb{Z}_2$-parity symmetry
\begin{eqnarray}\label{eq:spin-parity}
    \left[H,\hat{\cal P}\right]=0,\qquad\hat{\cal P}=\prod_{j=1}^{3N}Z_j,\qquad\hat{\cal P}^2=1.
\end{eqnarray}
This plays an important role in solving the Hamiltonian and subsequently analyzing the associated conserved charges.



\section{Solution}
\label{sec:spectrum}
From the structure of the Hamiltonian \eqref{eq:(6,3)-H}, it is clear that the usual JW transformation cannot map the system to a set of non-interacting free-fermions. Nonetheless, we now construct a particular map that exposes the underlying free-fermionic structure of the system concerned. To this end, consider the multi-site Majorana operator \footnote{In later parts of the paper we also use the notation $\mathcal{M}_j\equiv\mathcal{M}_{(3j-2)\cdots (3j+3)}$.}
\begin{eqnarray}\label{eq:multi-Majorana}
    {\cal M}_{j}(\Omega)=\left(\omega_1\gamma_{3j-2}+\omega_2\gamma_{3j-1}+\omega_3\gamma_{3j}\right)\times\nonumber\\\gamma_{3j+1}\gamma_{3j+2}\gamma_{3j+3},
\end{eqnarray}
with the normalization condition
\begin{eqnarray}
    \|\Omega\|^2=\omega_1^2+\omega_2^2+\omega_3^2=1.
\end{eqnarray}
Here we introduce the real vector $\Omega=(\omega_1~\omega_2~\omega_3)$. As can be verified, ${\cal M}_j$'s satisfy the relations
\begin{eqnarray}\label{eq:M-algebra}
    &{\cal M}_j(\Omega)^2=1,\qquad \{{\cal M}_j(\Omega),{\cal M}_{j+1}(\tilde{\Omega})\}=0,\nonumber\\
    &\left[{\cal M}_j(\Omega),{\cal M}_{k}(\tilde{\Omega})\right]=0,\quad\lvert j-k\rvert\geq 2.
\end{eqnarray}
The algebra generated by $\mathcal{M}_j(\Omega)$'s is isomorphic to {\it extraspecial 2-groups}. The latter has a center and commutator subgroup isomorphic to $\mathbb{Z}_2$ [See \cite{franko2006extraspecial} for further details]. This is generated by the operators $\prod_{j\in {\rm even}}\mathcal{M}_j$ and $\prod_{j\in {\rm odd}}\mathcal{M}_j$ as they commute with each generator $\mathcal{M}_j$.  For the standard TFIM, these are just the spin parity and the trivial identity operator for a closed chain. These operators are global discrete symmetries and are distinct from the local conserved charges to be introduced in Section \ref{subsec:FF-structure}.

We now define the local Majorana Hamiltonian
\begin{eqnarray}\label{eq:MajoranaIsingHam}
    H^{\cal F}(\Omega)=g\sum_{j=1}^{N}{\cal M}_{2j-1}(\Omega)+\sum_{j=1}^{N}{\cal M}_{2j}(\Omega),
\end{eqnarray}
with the identification $\gamma_{6N+j}\simeq \gamma_j$. In other words, $H^{\cal F}$ describes a Majorana chain on $6N$ Majoranas, with the local densities given by ${\cal M}_j$. Using the JW transformation \eqref{eq:JW1-transformation}, one can rewrite the local terms as
\begin{eqnarray}
    &&{\cal M}_{2j-1}(\Omega)=\omega_2 X_{3j-2}Y_{3j-1}Z_{3j}-\omega_1 Y_{3j-2}Y_{3j-1}Z_{3j}\nonumber\\&&\hspace{5cm}+\omega_3 Z_{3j-1}Z_{3j},\nonumber\\
    &&{\cal M}_{2j}(\Omega)=\omega_3 X_{3j}X_{3j+2}-\omega_2 Y_{3j}X_{3j+2}\nonumber\\&&\hspace{4cm}+\omega_1 X_{3j-1}Z_{3j}X_{3j+2}.
\end{eqnarray}
Strictly speaking, the above relations between the fermions and the spins hold when $j\neq N$. For $j=N$, the local fermionic boundary term is mapped to a non-local spin-1/2 operator. This plays an important role in determining the non-invertible symmetries of the standard TFIM \cite{Sinha:2025wqf}. We will address its consequences for this system in Section \ref{sec:integrability}.

We now demonstrate that the Majorana Hamiltonian $H^{\cal F}$ shares the same spectrum as that of the original Hamiltonian $H$ in \eqref{eq:(6,3)-H}. Indeed, setting $\omega_2=0$ yields our spin-Hamiltonian $H$ in \eqref{eq:(6,3)-H}, with the restrictions $\zeta=\tilde{\zeta},~\eta=-\tilde{\eta}$. However, as we show below, these apparent restrictions on the parameter space can be avoided by appropriate rotations of the local terms, with no need to impose $\omega_2=0$. Consider the parity preserving unitary operator ${\cal R}(\Omega)^\dagger{\cal R}(\Omega)=1$, which is a product of local unitaries
\begin{eqnarray}\label{eq:rotation}
    &&{R}(\Omega)=\prod_{j=1}^Ne^{-{\rm i}(\theta_{12}/2)Z_{3j-2}}e^{-{\rm i}(\theta_{23}/2)Z_{3j}},\nonumber\\
    &&\tan(\theta_{ij})=\frac{\omega_i}{\omega_j},\qquad [R(\Omega),\hat{\cal P}]=0.
\end{eqnarray}
It is not hard to verify that, it implements ${R}(\Omega){\cal M}_{2j-1(2j)}{R}(\Omega)^\dagger=h^{z(xx)}_j$, leading to
\begin{eqnarray}
    {R}(\Omega)H^{\cal F}(\Omega){R}(\Omega)^\dagger=H,
\end{eqnarray}
with the identifications
\begin{eqnarray}\label{eq:Bandomega}
    \eta=\left(\omega_1^2+\omega_2^2\right)^{\frac{1}{2}},~ {\zeta}=\omega_3,~\tilde{\eta}=\omega_1,~ \tilde{\zeta}=\left(\omega_2^2+\omega_3^2\right)^{\frac{1}{2}}.
\end{eqnarray}
As emphasized before, the above fermion-spin mapping is exact only in the bulk. We now proceed to solve the fermionic Hamiltonian ${H}^{\cal F}(\Omega)$, which in turn reveals the complete spectrum of the original Hamiltonian \eqref{eq:(6,3)-H}.   
\subsection{Free-fermionic structure}
\label{subsec:FF-structure}
As the neighboring terms in the Majorana Hamiltonian are shifted by three sites, we can further simplify this Hamiltonian by introducing the {\it effective Majoranas} through the redefinition,
\begin{eqnarray}\label{eq:MajtoMaj}
    &&a_j=\left(\omega_1\gamma_{3j-2}+\omega_2\gamma_{3j-1}+\omega_3\gamma_{3j}\right),\nonumber\\
    &&b_j=-{\rm i}\gamma_{3j-2}\gamma_{3j-1}\gamma_{3j}.
\end{eqnarray}
We introduce the notation $\Phi_{\rm Majorana}$ to symbolically represent the above mapping between the physical Majorana fermions $\gamma_j:j=1,\cdots,6N$ and the effective fermionic operators $a_j,b_j:j=1,\cdots,2N$. Note that this transformation is non-invertible. The operators $a_j,b_j$ are hermitian and are indeed effective Majoranas as they satisfy the relations
\begin{eqnarray}\label{eq:effective-Majorana-relations}
    &\left\{a_j,a_k\right\}=2\delta_{j,k}=\left\{b_j,b_k\right\},\nonumber\\
    &\left\{a_j,b_k\right\}=2\delta_{j,k}\hat{\cal C}_j,,\qquad\left[\hat{\cal C}_j,a_k\right]=0=\left[\hat{\cal C}_j,b_k\right],
\end{eqnarray}
for $\forall j,k$. Note that the $a_j,b_j$'s do not mutually anticommute on the same site forcing us to introduce the operators 
\begin{eqnarray}\label{eq:C-properties}
    \hat{\cal C}_j=\frac{1}{2}\left\{{a}_j,b_j\right\}={a}_jb_j={b}_j{a}_j,\qquad \hat{\cal C}_j^2=1.
\end{eqnarray}
With this mapping the multi-site Majorana operator ${\cal M}_j$ becomes
\begin{eqnarray}
    {\cal M}_j=\mathrm{i}~a_jb_{j+1},
\end{eqnarray}
and the corresponding Hamiltonian \eqref{eq:MajoranaIsingHam} can be expressed as bilinears in these effective Majoranas,
\begin{eqnarray}\label{eq:H-ff-form}
    H^{\cal F}={\rm i}g\sum_{j=1}^{N}{a}_{2j-1}{b}_{2j}+{\rm i}\sum_{j=1}^{N}{a}_{2j}{b}_{2j+1}.
\end{eqnarray}
From this point onward, we omit the explicit dependence on $\Omega$ and assume that condition \eqref{eq:Bandomega} is satisfied. One now can construct the fermionic parity $\hat{\cal P}$ in \eqref{eq:spin-parity} as
\begin{eqnarray}
    \hat{\cal P}={\rm i}^N\prod_{j=1}^{2N}b_j=\prod_{j=1}^{3N}Z_j,\qquad \left[H^{\cal F},\hat{\cal P}\right]=0.
\end{eqnarray}
Note that, the product $\prod_ja_j$ can be expressed using the charges $\{\hat{\cal C}_j\},\hat{\cal P}$ and hence is conserved as well.

The operator in \eqref{eq:H-ff-form} is still not free-fermionic due to the two flavors of the effective Majoranas in the Hamiltonian density. However, we get around this by noting that the operators $\hat{\cal C}_j$'s mutually commute $[\hat{\cal C}_j,\hat{\cal C}_k]=0$. In other words, from the relations \eqref{eq:effective-Majorana-relations}, the $\{\hat{\cal C}_j\}$'s are the central elements of the algebra generated by $a_j,b_j$.
This implies that the Hamiltonian $H^{\cal F}$ comes with an extensive number of strictly local conserved charges
\begin{eqnarray}
    [H^{\cal F},\hat{\cal C}_j]=0,\qquad \forall j.
\end{eqnarray}
Therefore, they have no dynamics under the time-evolution. Furthermore, since $\hat{\cal C}_j$ commutes with all ${a}_k,{b}_k$, the complete algebra of observables generated by the $\{{a}_k,{b}_k\}$ admits a block diagonal representation, labeled by the eigenvalues of $\hat{\cal C}_j$. Therefore, these central elements behaves more like a classical background field under which the Hamiltonian block diagonalizes\footnote{This structure is similar to what is observed in Kitaev's honeycomb model \cite{KITAEV20062}, where the Hamiltonian block diagonalizes in the common eigenspace of the set of local conserved gauge degrees of freedom. }. These are precisely the spin-1/2 operators \eqref{eq:spin-central}, rotated by the unitary $R(\Omega)$ in \eqref{eq:rotation} as
\begin{eqnarray}
    R(\Omega)\hat{\cal C}_{2j-1}R(\Omega)^{-1}={\cal I}_j,\quad R(\Omega)\hat{\cal C}_{2j}R(\Omega)^{-1}=\tilde{\cal I}_j.
\end{eqnarray}
We now discuss how these symmetries restrict the dynamics and break the total Hilbert space into several superselction sectors. As a result, the Hilbert space splits into $2^{2N}$ sectors as
\begin{eqnarray}
    {\cal H}=\bigoplus_{\{c_j\}}{\cal H}_{\{c_j\}},
\end{eqnarray}
where $c_j=\pm1$ are the eigenvalues of the operator $\hat{\cal C}_j$ and $\{c_j\}$ denotes the collective eigenvalues of the set $\{\hat{\cal C}_j\}$. Let us now see how these conserved charges help us in obtaining the spectrum of the Hamiltonian. Observe that, we can always write ${b}_j=\hat{\cal C}_j {a}_j$, yielding
\begin{eqnarray}
    H^{\cal F}={\rm i}g\sum_{j=1}^{N}{b}_{2j-1}\hat{\cal C}_{2j-1}{b}_{2j}+{\rm i}\sum_{j=1}^{N}{b}_{2j}\hat{\cal C}_{2j}{b}_{2j+1}.
\end{eqnarray}
Subsequently, the Hamiltonian simplifies in the sector ${\cal H}_{\{c_j\}}$ as
\begin{eqnarray}\label{eq:TFIM-b-Majorana}
    H^{\cal F}_{\{c_j\}}={\rm i}g\sum_{j=1}^{N}c_{2j-1}{b}_{2j-1}{b}_{2j}+{\rm i}\sum_{j=1}^{N}c_{2j}{b}_{2j}{b}_{2j+1},
\end{eqnarray}
which essentially describes quadratic Majorana chain with nearest-neighbor couplings given by $c_{j}=\pm 1$. Hence the spectrum of total Hamiltonian can be obtained by solving the quadratic Hamiltonians $H^{\cal F}_{\{c_j\}}$. This establishes that the seemingly interacting Hamiltonian $H$ in \eqref{eq:(6,3)-H} is actually free-fermionic in a single flavor of the effective Majorana.

\subsection{Hidden TFIM Hamiltonian}
\label{subsec:hiddenIsing}
\begin{figure}[H]
    \centering
    \begin{tikzpicture}

\node (A) at (0,0) {$\{\gamma_j|\,j=1,\cdots,6N\}$};

\node (B) at (5,0) {$\{X_j,Y_j,Z_j|\,j=1,\cdots,3N\}$};

\node (C) at (0,-2) {$\{a_j,b_j|\,j=1,\cdots,2N\}$};

\node (D) at (5,-2) {$\{\tilde{X}_j,\tilde{Y}_j,\tilde{Z}_j|\,j=1,\cdots,N\}$};

\draw[stealth - , thick] (A) -- (B) node[midway, above] {$\Phi_{\rm JW_1}$};
\draw[-stealth, thick] (A) -- (C) node[midway, left] {$\Phi_{\rm Majorana}$};
\draw[-stealth, thick] (C) -- (D) node[midway, below] {$\Phi_{\rm JW_2}$};
\draw[-stealth, thick] (B) -- (D) node[midway, right] {$\Phi_{\rm Spin}$};

\end{tikzpicture}
    \caption{A commutative diagram explaining various transformations between fermions and spins. The mapping $\Phi_{\rm Spin}$ can be obtained by composing $\Phi_{{\rm JW}_1}$ \eqref{eq:JW1-transformation}, $\Phi_{\rm Majorana}$ \eqref{eq:MajtoMaj} and $\Phi_{{\rm JW}_2}$ \eqref{eq:JW2-transform}.}
    \label{fig:schematicMap}
\end{figure}
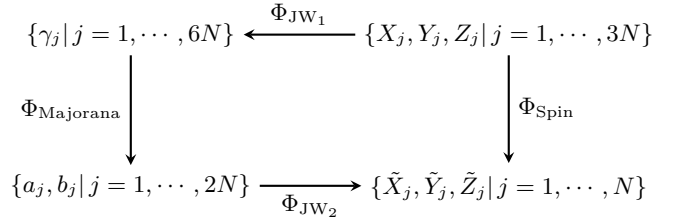

Before solving for the spectrum, we will summarize the steps leading to the TFIM Hamiltonian on a reduced number of lattice sites. Let us introduce a different JW transformation 
\begin{eqnarray}\label{eq:JW2-transform}
    b_{2j-1}=\left(\prod_{k=1}^{j-1}\tilde{Z}_k\right)\tilde{X}_j,\qquad b_{2j}=\left(\prod_{k=1}^{j-1}\tilde{Z}_k\right)\tilde{Y}_j,
\end{eqnarray}
with $j=1,\cdots,N$, which maps the effective Majoranas $b_j$'s to half the number of spin variables $\tilde{X}$, $\tilde{Y}$ and $\tilde{Z}$.
We will symbolically refer to this mapping as $\Phi_{{\rm JW}_2}$. On applying this trasnformation on the Hamiltonian, \eqref{eq:TFIM-b-Majorana} we obtain the standard TFIM Hamiltonian
\begin{eqnarray}\label{eq:Hidden_Ising}
    H^{\cal F}_{\{c_j\}}=-g\sum_{j}c_{2j-1}\tilde{Z}_j-\sum_jc_{2j}\tilde{X}_j\tilde{X}_{j+1},
\end{eqnarray}
acting on a $2^N$ dimensional Hilbert space. This establishes that the original $3N$-site Hamiltonian in \eqref{eq:(6,3)-H} decomposes as a direct sum of several quantum Ising chains on a reduced $N$-number of effective sites. On each sector, the coupling coefficients of the different local Hamiltonian densities are determined by the eigenvalues of the central elements $\{c_j\}$. These steps are summarized in the commutative diagram Figure \ref{fig:schematicMap}.

It should be noted that, the transformation
\begin{eqnarray}
    \Phi_{\rm Spin}={\Phi}_{{\rm JW}_2}\circ\Phi_{\rm Majorana}\circ\Phi_{{\rm JW}_1}
\end{eqnarray}
maps the original Pauli matrices $X_j,Y_j,Z_J$ to the effective spin-$1/2$ operators $\tilde{X}_j,\tilde{Y}_j,\tilde{Z}_j$, thus making explicit the connection between the original $3N$-site spin chain and the corresponding effective Ising chain defined over $N$ sites. We remark that the mapping in \eqref{eq:MajtoMaj} is by no means the unique realization of the algebra \eqref{eq:effective-Majorana-relations}. In Section \ref{sec:gYBE-HFF}, we shall present several explicit mappings from $\{\gamma_j\}$ to $\{a_j,b_j\}$, all of which furnish valid realizations of \eqref{eq:effective-Majorana-relations}. Consequently, all such models possess an underlying hidden Ising structure.

\subsection{Spectrum}
\label{subsec:spectrum}
Having established that the Hamiltonian \eqref{eq:(6,3)-H} is free-fermionic for arbitrary values of $g$, we now specialize to the case $g=1$. As will be shown in Section~\ref{sec:integrability}, this particular Hamiltonian can be derived from a generalized $R$-matrix. We shall discuss both the closed and the open boundary conditions. Interestingly, when considering a closed chain, imposing either the periodic or the anti-periodic boundary conditions yield the same result. It is straightforward to extend this analysis for arbitrary $g$.

\begin{figure*}[t]
\centering

\includegraphics[height=0.3\textwidth]{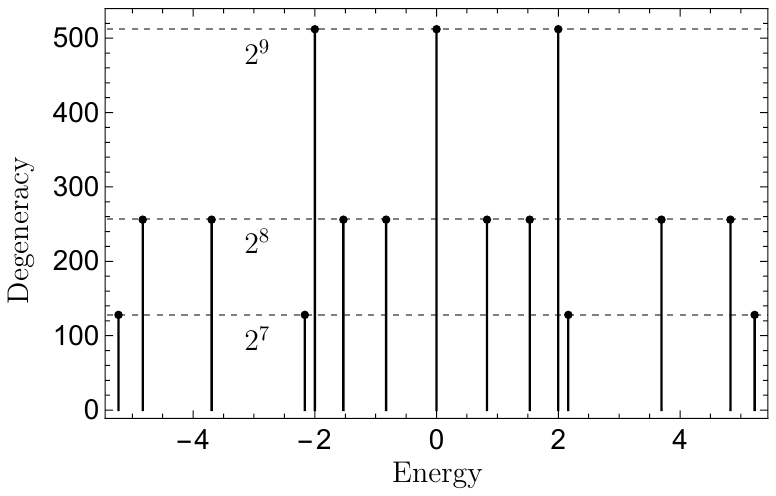}
\hfill
\includegraphics[height=0.3\textwidth]{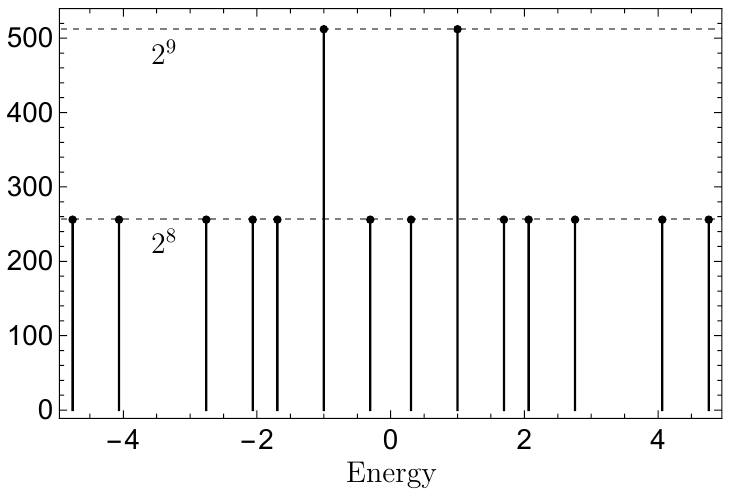}

\caption{
(a) Periodic boundary conditions and (b) open boundary conditions.
The energy eigenvalues of $H_+$ and the corresponding degeneracies for $N=4$.
The doubling of degeneracy is evident.
}
\label{fig:placeholder}
\end{figure*}

\subsubsection{Closed chain}
\label{subsubsec:closed-chain-spectrum}
Consider the fermionic Hamiltonians
\begin{eqnarray}\label{eq:PBC_APBC}
    H_{\pm}={\rm i}\sum_{j=1}^{2N-1}{a}_j{b}_{j+1} \pm{\rm i}{a}_{2N}{b}_{1},
\end{eqnarray}
where $\pm$ denote periodic and antiperiodic boundary conditions, respectively. However, it turns out that both $H_\pm$ share the same spectrum. To see this, consider the operator $\gamma^\varepsilon$ defined as
\begin{eqnarray}
   \gamma^\varepsilon=\sum_{k=1}^3\varepsilon_k\gamma_k,&\quad& \sum_{k=1}^3\varepsilon_k\omega_k=0,\nonumber\\
   \gamma^\varepsilon a_j=-a_j\gamma^\varepsilon,\quad&&\gamma^\varepsilon b_j=-(-1)^{\delta_{j,1}}b_j\gamma^\varepsilon,
\end{eqnarray}
with $j=1,\cdots,2N$. It is now easy to verify that
\begin{eqnarray}\label{eq:PBC_APBC_Equiv}
    \left(\gamma^\varepsilon\right) H_\pm\left(\gamma^\varepsilon\right)^{-1}=H_\mp,
\end{eqnarray}
establishing that $H_\pm$ essentially have the same spectrum. This plays a crucial role in determining the symmetries of the Hamiltonian \eqref{eq:(6,3)-H}. 

Therefore, we may restrict our analysis to periodic boundary conditions without any loss of generality. The corresponding Hamiltonian in the sector $\{c_j\}$ takes the form
\begin{eqnarray}\label{eq:H_c}
    H_{\{c_j\}}={\rm i}\sum_{j=1}^{2N}c_{j}{b}_j{b}_{j+1},
\end{eqnarray}
Although there are $2^{2N}$ different possibilities for $H_{\{c_j\}}$, as far as the energy eigenvalues are concerned, effectively there are only two inequivalent sectors. To see this, consider 
\begin{eqnarray}
    {\frak H}_+={\rm i}\sum_{j=1}^{2N}{b}_j{b}_{j+1},\qquad U_{kl}=\prod_{r=k+1}^l{b}_r.
\end{eqnarray}
Evidently, one obtains ${\frak H}_+$ by setting $c_j=1$ for every $j$ in $H_+$ \eqref{eq:H_c}. The operator $U_{kl}$ is a nonlocal Majorana string, having support over all the sites between $k+1$ and $l$. It is then straightforward to check that
\begin{eqnarray}
    U_{kl}{\frak H}_+U_{kl}^{-1}={\rm i}\sum_{j\neq k,l}{b}_{j}{b}_{j+1} -{\rm i}\left({b}_k{b}_{k+1}+{b}_l{b}_{l+1}\right).
\end{eqnarray}
In other words, all the Hamiltonians $H_{\{c_j\}}$ that differ from ${\frak H}_+$ on even number of sites are equivalent to ${\frak H}_+$ and have the same spectrum as that of ${\frak H}_+$. The number of such isospectral sectors is readily given by
\begin{eqnarray}
    \sum_{k=0}^{N}\binom{2N}{2k}=2^{2N-1}.
\end{eqnarray}
This closely resembles the Hilbert space structure of the FFDs, as detailed in \cite{vernier2026hilbert}. Similarly, the rest $2^{2N}-2^{2N-1}=2^{2N-1}$ possible Hamiltonians $H_{\{c_j\}}$ which differ from ${\frak H}_+$ on odd number of sites, are equivalent to
\begin{eqnarray}
   {\frak H}_-={\rm i}\sum_{j=1}^{2N-1}{b}_j{b}_{j+1} -{\rm i}\,{b}_{2N}{b}_{1}.
\end{eqnarray}
Therefore, knowing the spectra of ${\frak H}_\pm$ suffices to solve the quartic Hamiltonian \eqref{eq:(6,3)-H}. Physically, ${\frak H}_\pm$ describe free Majorana chains with periodic and anti-periodic boundary conditions, respectively. They can be diagonalized in the momentum space as 
\begin{eqnarray}
    {\frak H}_\pm=\sum_{n_\pm} {\cal E}_\pm(n_\pm)d_{n_\pm}^\dagger d_{n_\pm}+\text{constant},
\end{eqnarray}
where $d_{n_\pm}$ are the momentum modes. The single-particle energies are given by 
\begin{eqnarray}
    {\cal E}_\pm(n_\pm)=4\sin\left(\frac{\pi n_\pm}{N}\right),
\end{eqnarray}
with $n_+=0,1,\cdots,N-1,N$ and $n_-=\frac{1}{2},\cdots,N-\frac{1}{2}$. Each energy level is at least $2^{2N-1}$-fold degenerate. Furthermore, the spectrum is gapless as well.

Interestingly, one can anticipate the above results from a different perspective. Notice that two Hamiltonians $H_{\{c_j\}}$ and $H_{\{c'_j\}}$ are equivalent if $c'_j=\theta_jc_j\theta_{j+1}$, where $\theta_j=\pm 1$. This directly follows from the fact that under the transformation $b_j\to \theta_jb_j$, the Majorana algebra $\{b_j,b_k\}=2\delta_{j,k}$ remains unchanged and hence describes the same physical situation. Let us now consider 
\begin{eqnarray}
    {\cal W}=\prod_{j=1}^{2N}\hat{\cal C}_j,\qquad {\cal W}^2=1.
\end{eqnarray}
Then the condition for equivalence between two different Hamiltonians translates into 
\begin{eqnarray}
    {\cal W}_{\{c'_j\}}=\prod_{j=1}^{2N}c'_j=\prod_{j=1}^{2N}(\theta_jc_j\theta_{j+1})={\cal W}_{\{c_j\}},
\end{eqnarray}
where we used the fact that $\theta_j^2=1$. In other words, ${\cal W}$ is an invariant for the equivalent Hamiltonians. Since ${\cal W}^2=1$, there exist only two scenarios ${\cal W}_{\{c_j\}}=\pm 1$, which precisely correspond to the cases with even and odd number of $c_j=-1$, respectively. This is exactly what we obtained rigorously using the Majorana strings $U_{kl}$'s. The operator ${\cal W}$ can be thought of as some gauge-invariant Wilson loop.

\subsubsection{Open chain}
\label{subsubsec:open-chain-spectrum}
We now briefly talk about the open boundary condition, for which the Hamiltonian becomes
\begin{eqnarray}\label{eq:openH}
    H_{\rm o}={\rm i}\sum_{j=1}^{2N-1}{a}_j{b}_{j+1}={\rm i}\sum_{j=1}^{2N-1}{b}_j\hat{\cal C}_{j}{b}_{j+1}.
\end{eqnarray}
In this case, the degeneracy is enhanced further and becomes twice as large as only a single sector admits non-trivial spectrum. The boundary modes ${b}_{2N},{b}_{1}$, along with the operators $U_{kl}$, suffice to establish the spectral equivalence of all the sectors. The resulting degeneracy is therefore $2^{2N}$, rather than $2^{2N-1}$. To obtain the spectrum, we set $c_j=1$ in \eqref{eq:openH} for all $j$ with
\begin{eqnarray}
    {\frak H}_{\rm o}={\rm i}\sum_{j=1}^{2N-1}{b}_j{b}_{j+1}=\sum_{n_{\rm o}}{\cal E}_{\rm o}(n_{\rm o})d^\dagger_{n_{\rm o}}d_{n_{\rm o}}+{\rm constant}.
\end{eqnarray}
The single-particle dispersion relation is given by
\begin{eqnarray}\label{eq:spectrum-open}
    {\cal E}_{\rm o}(n_{\rm o})=4\sin\left(\frac{2\pi n_{\rm o}}{2N+1}\right),\qquad n_{\rm o}=1,\cdots,N.
\end{eqnarray}
In contrast to the closed boundary case, each energy levels are $2^{2N}$-fold degenerate. This also can be understood from the fact that for open boundaries, the equivalence between two Hamiltonians $H_{\{c_j\}}$ and $H_{\{c'_j\}}$ does not impose any constraint relating $c_{2N}$ and $c'_{2N}$. This freedom can be exploited to show that there is only one inequivalent spectrum in this case.

\section{Thermodynamics of the model}
\label{sec:thermodynamics}
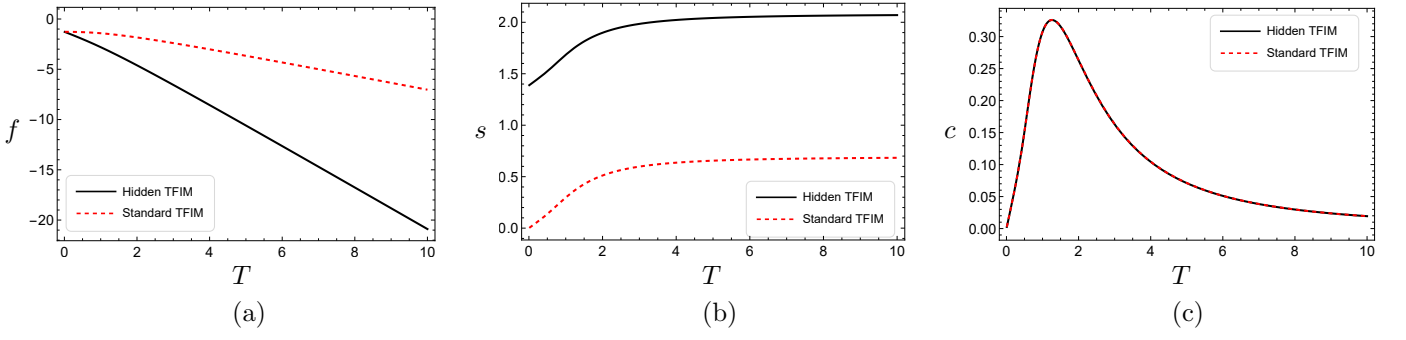
\begin{figure*}[t]
\centering

\begin{overpic}[width=0.3\textwidth]{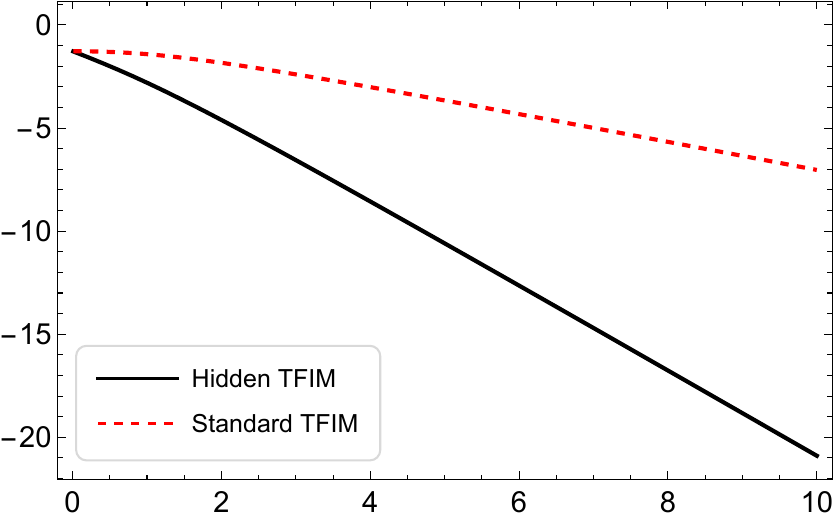}
    \put(-6,30){$f$}
    \put(50,-6){$T$}
    \put(50,-15){(a)}
\end{overpic}
\hfill
\begin{overpic}[width=0.3\textwidth]{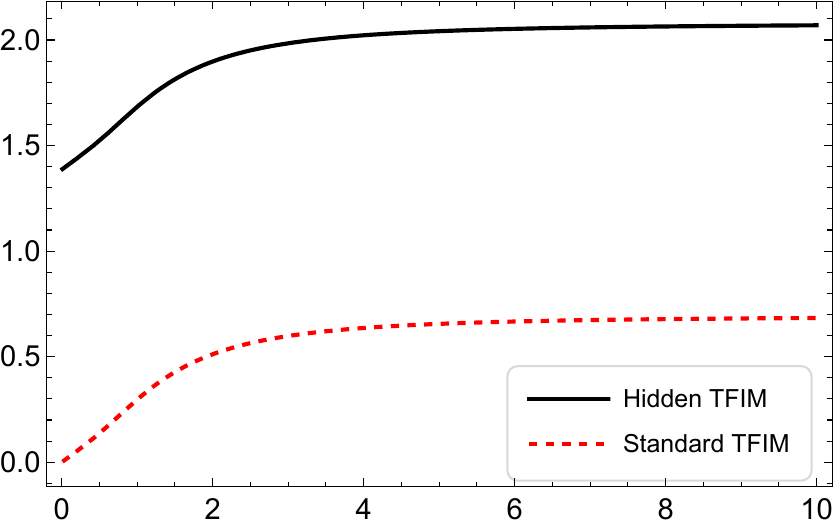}
    \put(-6,30){$s$}
    \put(50,-6){$T$}
    \put(50,-15){(b)}
\end{overpic}
\hfill
\begin{overpic}[width=0.3\textwidth]{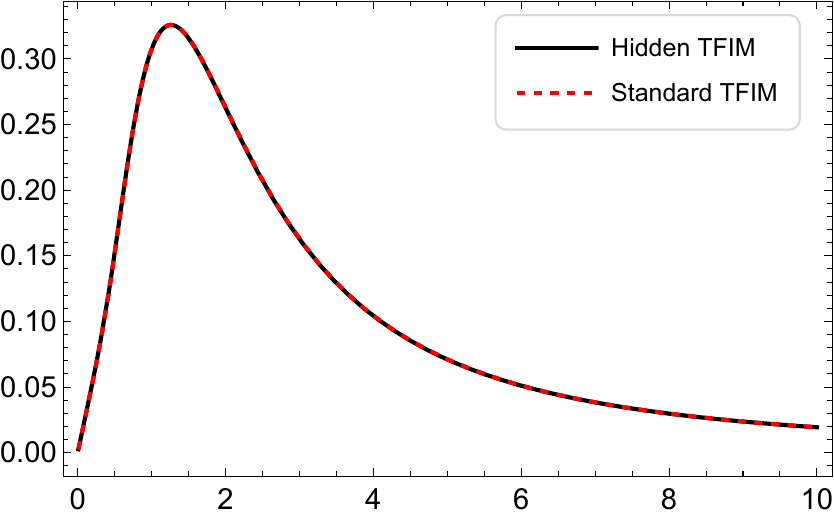}
    \put(-6,30){$c$}
    \put(50,-6){$T$}
    \put(50,-15){(c)}
\end{overpic}
\vspace{1cm}
\caption{
Thermodynamic properties of the hidden critical ($g=1$) transverse-field quantum Ising model for a system size $N=200$, with $k_B=1$ in natural units, are shown as functions of temperature $T$: (a) free-energy density $f(T)$, (b) thermal entropy per site $s(T)$, and (c) heat capacity per site $c(T)$. The results are obtained using the energy spectrum given in Eq.~\eqref{eq:spectrum-open}.
}
\label{fig:Thermo-plots}

\end{figure*}
We will now evaluate the thermal quantities associated to this spin chain \cite{Takahashi_1999}. Start with the explicit form of the energy,
\begin{equation}
    E = \sum_n \mathcal{E}(n) m_n + E_0,   
\end{equation}
where $m_n \in \{0,1\}$ denotes the occupation number for $n^{\rm th}$ fermionic mode, and  $E_0 = - \frac{1}{2}\sum_n \mathcal{E}(n)$  specifies the ground state energy for the system. The finite temperature canonical partition function can be written as
\begin{eqnarray}
    \mathcal{Z}_{\rm open} &=&2^{2N}\sum\limits_{\{m_{n_{\rm o}}\}} \exp\left[{-\beta\left( E_0+\sum_n \mathcal{E}_{\rm o}(n_{\rm o})m_{n_{\rm o}}\right)}\right]\nonumber\\
    &=& 2^{2N}\prod_{n_{\rm o}}2 \cosh{\frac{\beta \mathcal{E}_{\rm o}(n_{\rm o})}{2}}.
\end{eqnarray}
This is similar to the partition function of the TFIM modulo the multiplicative degeneracy factor, $2^{2N}$. The temperature and the Boltzmann factor are $T$ and $k_B$ respectively. Note that the partition function considered above is valid for an open chain, while the closed-chain case differs slightly due to boundary conditions. As previously discussed, in the case of closed-chain, the Hilbert space decomposes into two inequivalent sectors, each described by equal degeneracies of $2^{2N-1}$. Therefore, the partition function can be described as 
\begin{eqnarray}
     &&\mathcal{Z}_{\rm closed}\nonumber\\
     &=&2^{2N-1}\left[\prod_{n_{+}}2 \cosh{\frac{\beta \mathcal{E}_+({n_{+}})}{2}}+\prod_{n_{-}} \cosh{\frac{\beta \mathcal{E}_-({n_{-}})}{2}}\right].\nonumber\\
\end{eqnarray}
It is evident from the spectrum at the critical point $g=1$ that, in the large $N$ limit, the effect of boundary conditions becomes negligible as the quantization conditions become irrelevant. Consequently, the two products corresponding to $n_+$ and $n_-$ become identical in this thermodynamic limit. Therefore, the partition function for the closed chain reduces to the same expression as that obtained for the open chain. Strictly speaking, however, for a finite number of sites, the partition function of the closed chain remains distinct from that of the open case, as is clearly evident from the expressions given above.

We want to compute the thermodynamic quantities in thermodynamic limit, so it is sufficient to work with the partition function for the open chain. First, we note that the control parameters for this system are the temperature $T$, and the parameter $g$ in the Hamiltonian, that can be related to the external magnetic field. This is also evident from the expressions for the partition function. Note that the energy eigenvalues, $\mathcal{E}_o(n_o)$ are in general a function $g$, though we have only computed the spectrum for $g=1$ in Section \ref{subsec:spectrum}. We will see that this does not make a difference to the results below   Therefore, the expression of the free energy is given by
\begin{widetext}
\begin{eqnarray}
    F=-\frac{1}{\beta}\ln{\mathcal{Z}_{\rm open} }=-\frac{1}{\beta}\left[2N \ln{2}+\sum_n \ln{2 \cosh{\frac{\beta \mathcal{E}(n)}{2}}}\right].
\end{eqnarray}
The extra factor can be attributed to the exponential degeneracy of the energy levels of this system. In the thermodynamic limit, the sum will be replaced by an integral of the form $\sum_n \to \int_{-N/2}^{N/2}{\rm d}n$. Subsequently, we can express the free energy density as 
    \begin{equation}
    f(T)=\frac{F}{N} = -3k_BT\ln{2}- \frac{2k_BT}{N}\int\limits_{0}^{N/2} \ln{\cosh{\frac{\beta \mathcal{E}(n)}{2}}}{\rm d}n.
\end{equation}
The thermal entropy and heat capacity per site are expressed as  
\begin{eqnarray}
    s(T)&=&\frac{S}{N}= -\frac{1}{N}\frac{\partial F}{\partial T}= 3k_B \ln{2}+\frac{2k_B}{N}\int\limits_{0}^{N/2} \left(\ln{\cosh{\frac{\beta \mathcal{E}(n)}{2}}}-\frac{\beta\mathcal{E}(n)}{2}\tanh{\frac{\beta \mathcal{E}(n)}{2}}\right){\rm d}n,\nonumber\\
    c(T)&=&\frac{C}{N}=-\frac{T}{N}\frac{\partial^2 F}{\partial T^2}=\frac{2k_B}{N}\int\limits_{0}^{N/2} \left(\frac{\beta \mathcal{E}(n)}{2}\right)^2 \mathrm{sech}^2{\frac{\beta \mathcal{E}(n)}{2}}{\rm d}n.
\end{eqnarray}
\end{widetext}

All thermodynamic quantities computed above, are plotted against $T$ in Figures \ref{fig:Thermo-plots}.
It is observed that  at low temperature limit, the free energy converges to the ground state energy and in case of high temperatures, it has a linear $T$ dependence as expected from the usual TFIM model. It is also noted that as temperature goes to infinity, the value of thermal entropy per site turns out to be  $s(\infty) = s_{\mathrm{standard}} +2\ln{2}$, indicating higher number of free states in high temperatures. On the other hand, at low-temperature limit, the hidden Ising model retains residual thermal entropy density $s(0) = 2\ln{2}$, which implies a massive degeneracy in the ground state that scales exponentially with the system size, whereas in the thermodynamic limit for usual TFIM, the thermal entropy density reduces to zero as per the \textit{Third Law of Thermodynamics}.

The next natural quantity to compute are the static and dynamical correlation functions of this model. However, this is an involved computation for the standard TFIM as well, and it is known to be associated with many interesting mathematical structures, such as Toda-type differential equations, Painlev\'{e} transcendents and Fredholm and Toeplitz determinants \cite{Barouch1971StatisticalMO,MCCOY198335,McCoy1993LatticeMI,Mahoney_2024,Perk_2009, Essler_2016}. The correlation functions other closely related free-fermionic systems like the $XX$-spin chain are known to be related to the Ablowitz-Ladik PDE \cite{Its_1993}. The latter is also known to be equivalent to the discrete non-linear Schr\"{o}dinger equation \cite{Hoffmann_2000}. See also \cite{Its-DE-correlations} and Chapter 6 os \cite{Korepin1993QuantumIS} for more details on differential equations for correlation functions. 

We postpone the investigation of the static and dynamic correlation functions for the hidden Ising case as they are both analytically and numerically quite demanding. The source of the numerical difficulties is the fact that we are allowed to choose lattices with sizes in multiples of 3 only. Our numerical resources prevent us from then considering large $N$, and thus we are unable to make any meaningful statements about the behavior of these functions at present.

\section{Integrability and conserved charges}
\label{sec:integrability}



We will now show that the Hamiltonian in \eqref{eq:(6,3)-H} can be derived from a $R$-matrix that satisfies a multi-site generalization of the Yang-Baxter equation. To construct such multi-indexed $R$-matrices we will use the $\mathcal{M}$ operators introduced in \eqref{eq:M-algebra}, namely 
\begin{eqnarray}
    &{\cal M}_j^2=1,\qquad\{{\cal M}_j,{\cal M}_{j+1}\}=0,\nonumber\\
    &[{\cal M}_j,{\cal M}_k]=0,\qquad |j-k|\geq 2.
\end{eqnarray}
As remarked in Section \ref{sec:spectrum}, these operators generate an extraspecial 2-group. It should be emphasized that although ${\cal M}_j$ carries a single index, it actually is a multi-site operator, acting non-trivially on $6$ Majorana indices
\begin{eqnarray}
    &&{\cal M}_{(3j-2)\cdots(3j+3)}\equiv{\cal M}_j
    =\left(\omega_1\gamma_{3j-2}+\omega_2\gamma_{3j-1}+\omega_3\gamma_{3j}\right)\nonumber\\&&\hspace{4.5cm}\gamma_{3j+1}\gamma_{3j+2}\gamma_{3j+3}.
\end{eqnarray}
We now introduce the generalized $R$-matrix supported non-trivially on 6 indices as \footnote{For a proof of this solution and its generalization to an arbitrary number of indices, see Appendix \ref{app:gYBE}. }
\begin{eqnarray}\label{eq:(6,3)-R-1+3}
    R_{(3j-2)\cdots(3j+3)}(\lambda)=\mathbb{1} + \tan({\lambda})\mathcal{M}_{(3j-2)\cdots(3j+3)}.
\end{eqnarray}
We suppress the $\Omega$ parameter of $\mathcal{M}$ to highlight the spectral parameter, $\lambda$ dependence of the generalized $R$-matrix. The spectral parameter is taken to be a complex number. 
This can be seen as the {\it Baxterized} version of the constant braid operator\footnote{The algebraic realization of this operator can be understood as a generalization of the two-site braid operator in the $H0,2$ family, introduced in \cite{maity2024algebraic}.} $\sigma=\frac{\mathbb{1}+\mathrm{i}\,\mathcal{M}}{\sqrt{2}}$ \cite{rowell2010extraspecialtwogroupsgeneralizedyangbaxter}, with the $\mathrm{i}\mathcal{M}$ being generators of the extraspecial 2-group. The transfer matrix constructed out of such $R$-matrices were also shown to contain the clean versions of the SYK model \cite{hosho-SYK}. 

The $R$-matrix in \eqref{eq:(6,3)-R-1+3}, satisfies the spectral parameter dependent $(d,6,3)$-gYBE\footnote{This is a special case of the family of multi-site generalizations of the YBE denoted $(d,l,m)$-gYBE \cite{rowell2010extraspecialtwogroupsgeneralizedyangbaxter}. A short introduction is given in Appendix \ref{app:gYBE}.},
\begin{eqnarray}\label{eq:6-3-gYBE-braided}
   &&R_{1\cdots6}(\lambda)R_{4\cdots9}(\lambda+\mu)R_{1\cdots6}(\mu)\nonumber\\&&\qquad\qquad=R_{4\cdots9}(\mu)R_{1\cdots6}(\lambda+\mu)R_{4\cdots9}(\lambda).
\end{eqnarray}
$\lambda$, $\mu$ are complex spectral parameters. This is to be viewed as the braided form of a multi-site generalization of the YBE. For notational convenience, we now introduce the general operator
\begin{eqnarray}
    &&R_{\vec{j},\vec{k}}(\lambda)=\mathbb{1}+\tan(\lambda)\left(\omega_1\gamma_{3j-2}+\omega_2\gamma_{3j-1}+\omega_3\gamma_{3j}\right)\nonumber\\&&\hspace{5cm}\gamma_{3k-2}\gamma_{3k-1}\gamma_{3k},
\end{eqnarray}
with the shorthand vector notation
\begin{eqnarray}
    \vec{j}=(3j-2,3j-1,3j).
\end{eqnarray}
One immediately identifies that $R_{\vec{j},\vec{j+1}}=R_{(3j-2)\cdots(3j+3)}$. This can be used to transition to a non-braided form, that will be useful to construct integrable models as per the framework of the QISM. This is obtained by multiplying the generalized braided $R$-matrix with a multi-site permutation operator 
\begin{eqnarray}\label{eq:non-braided-R-6-site}
    &&\tilde{R}_{\vec{j},\vec{k}}(\lambda) = P^-_{3j-2,3k-2}P^-_{3j-1,3k-1}P^-_{3j,3k}R_{\vec{j},\vec{k}}(\lambda),\nonumber\\
    &&\qquad P^-_{j,k}=\frac{\gamma_{j}-\gamma_{k}}{\sqrt{2}},\qquad \left(P^-_{j,k}\right)^2=1.
\end{eqnarray}
being the permutation operator on the space of Majoranas \cite{Sinha:2025wqf}. It is now easy to see that $\tilde{R}_{\vec{j},\vec{k}}$ satisfies a non-braided version of the $(d,6,3)$-gYBE \eqref{eq:6-3-gYBE-braided} as
\begin{eqnarray}\label{eq:6-3-gYBE-non-braided}
   & & \tilde{R}_{\vec{1},\vec{2}}(\lambda)\tilde{R}_{\vec{1},\vec{3}}(\lambda+\mu)\tilde{R}_{\vec{2},\vec{3}}(\mu) \nonumber\\
    &&\qquad\qquad=\tilde{R}_{\vec{2},\vec{3}}(\mu)\tilde{R}_{\vec{1},\vec{3}}(\lambda+\mu)\tilde{R}_{\vec{1},\vec{2}}(\lambda).
\end{eqnarray}
The $R$-matrix \eqref{eq:non-braided-R-6-site} satisfies the following $RTT$-relation 
\begin{eqnarray}\label{eq:RTT}
    &&\tilde{R}_{\vec{\alpha},\vec{\beta}}(\lambda-\mu) T_{\vec{\alpha}}(\lambda)T_{\vec{\beta}}(\mu)\nonumber\\
    &&\qquad\qquad= T_{\vec{\beta}}(\mu) T_{\vec{\alpha}}(\lambda) \tilde{R}_{\vec{\alpha},\vec{\beta}}(\lambda-\mu),
\end{eqnarray}
where the monodromy matrix $T(\lambda)$ is defined as 
\begin{equation}\label{eq:monodromy}
    T_{\vec{\alpha}}(\lambda)=\tilde{R}_{\vec{\alpha},\vec{N}}(\lambda) \cdots \tilde{R}_{\vec{\alpha},\vec{1}}(\lambda).
\end{equation}
The $RTT$ relation is supported on six auxiliary indices, denoted as $\vec{\alpha}=(\alpha_1,\alpha_2,\alpha_3),\vec{\beta}=(\beta_1,\beta_2,\beta_3)$, with each monodromy matrix associated with three auxiliary indices. Note that, by this definition of the monodromy matrix, it is evident that the $T$-matrix can be defined provided that the total number of sites is an integer multiple of $3$. Deriving the Hamiltonian and the other conserved quantities from the $RTT$-relation \eqref{eq:RTT} involves several subtleties due to the anticommutating nature of the Majorana fermions. As discussed in detail in Appendix \ref{app:derivation-Hamiltonian}, the $RTT$-relation leads to the commuting transfer matrices
\begin{eqnarray}\label{eq:transfermatrix}
    &[\tau(\lambda),\tau(\mu)]=0,\nonumber\\
    &\tau(\lambda)={\rm tr}_{\vec{\alpha},\vec{\beta}} \left[T_{\vec{\alpha}}(\lambda)\right]={\rm tr}_{\vec{\alpha},\vec{\beta}} \left[T_{\vec{\beta}}(\lambda)\right],
\end{eqnarray}
where ${\rm tr}_{\vec{\alpha},\vec{\beta}}$ denotes partial trace over the auxiliary indices. The transfer matrix $\tau(\lambda)$ further gives rise to the mutually commuting local conserved charges, obtained by taking the logarithmic derivatives
\begin{eqnarray}
    \frac{{\rm d^r}}{{\rm d}\lambda^r}\log\tau(\lambda)\big|_{\lambda=0}.
\end{eqnarray}
Usually the first-order charge and the transfer matrix at $\lambda=0$ are associated with the local Hamiltonian and the appropriate translation operator, respectively. As shown in Appendix \ref{app:derivation-Hamiltonian}, the transfer matrix $\tau(\lambda)$ leads to the following expression for the Hamiltonian:
\begin{eqnarray}
   H_-&=&\frac{{\rm d}}{{\rm d}\lambda}\log\tau(\lambda)\big|_{\lambda=0}=\sum\limits_{j=1}^{2N}(-1)^{\delta_{j,2N}}{\cal M}_j.
\end{eqnarray}
We can now identify $H_-$ as the antiperiodic Hamiltonian introduced in \eqref{eq:PBC_APBC}. The corresponding translation operator becomes
\begin{eqnarray}\label{eq:translationAPBC}
    \tau(0)=\gamma_1\gamma_2\gamma_3\prod_{j=1}^{6N-3}\left(\frac{\gamma_j-\gamma_{j+3}}{\sqrt{2}}\right).
\end{eqnarray}
The higher-order charges can, in principle, be derived by repeatedly differentiating with respect to the spectral parameter $\lambda$. Nevertheless, there is a more straightforward procedure that enables the generation of these higher-order charges in a considerably simpler way.



\subsection{Boost operator and higher conserved charges}
\label{sec:Boost and higher conserved}
In the framework of QISM, the transfer matrix generates the set of conserved quantities and dictates the system's dynamics. As mentioned earlier, the logarithmic derivatives of the transfer matrix give rise to a tower of mutually commuting local conserved charges. However, instead of repeatedly performing the derivatives, the conserved charges can be obtained using the \textit{boost operator method} \cite{sogo1983boost,loebbert2016lectures}. We will demonstrate this relationship for our $R$-matrix \eqref{eq:non-braided-R-6-site} through the $(d,6,3)$-gYBE. The extensive derivation can be found in Appendix \ref{app:Boost-derivation}.  

To be precise, we shall find the conserved quantities for an infinite chain. The key ingredient is the \textit{boost operator} ${\cal B}$, defined as
\begin{eqnarray}
    {\cal B}=\sum_{j=-\infty}^\infty j{\cal M}_{j}.
\end{eqnarray}
As shown in Appendix \ref{app:Boost-derivation}, the transfer matrix and the boost operator satisfy the relation
\begin{eqnarray}
    \frac{{\rm d}\tau(\lambda)}{{\rm d}\lambda}=[{\cal B},\tau(\lambda)],\quad \tau(\lambda)=\tau(0)~ \exp\left[{\sum_{r=1}^{\infty}\lambda^r I_{r+1}}\right],
\end{eqnarray}
which in turn leads to the recursive relation 
\begin{eqnarray}\label{eq:Ifromboost}
    I_{r+1}=\frac{1}{r}[{\cal B},I_r],\qquad I_1:=\tau(0).
\end{eqnarray}
It can be checked that the operator $I_1=\tau(0)$ \eqref{eq:translationAPBC} behaves as a translation operator $I_1^{-1}\mathcal{M}_{j}I_1=\mathcal{M}_{j-1}$, as expected. Using this translational invariance, we can find out the next order conserved quantity (first-order charge) from the expression \eqref{eq:boost}
\begin{equation}
    I_2 = I_1^{-1}\mathcal{B}I_1-\mathcal{B}=\sum\limits_j \mathcal{M}_{j} =\mathrm{i} \sum_j a_j b_{j+1},
\end{equation}
which we can identify to be the Hamiltonian on an infinite chain. Similarly, from \eqref{eq:Ifromboost}, the next two conserved quantities can be obtained as 
\begin{eqnarray}
    &&I_3 =-\sum\limits_j \mathcal{M}_{j}\mathcal{M}_{j+1}={\rm i}\sum_j a_j \hat{\mathcal{C}}_{j+1} b_{j+2},\\
    &&I_4 =-\sum\limits_j \mathcal{M}_{j}\mathcal{M}_{j+1}\mathcal{M}_{j+2}={\rm i}\sum_j a_j \hat{\mathcal{C}}_{j+1}\hat{\mathcal{C}}_{j+2} b_{j+3}.\nonumber
\end{eqnarray}
The general expression of the conserved charge \footnote{The conserved charges of the standard one-dimensional Ising chain in a transverse magnetic field were first introduced in \cite{grady1982infinite}, where they were shown to be inherited from the conserved quantities of the $XYZ$-spin chain. Moreover, a general prescription based on a free-fermionic algebraic framework was recently presented in \cite{minami2025conserved}, in which the conserved charges are obtained from the kernel structure of a representation matrix, yielding the string-type local density operators.} can be written as 
\begin{equation}\label{eq:c-charges}
    I_{r+1} \simeq \sum \limits_j \mathcal{M}_j \cdots \mathcal{M}_{j+r-1}, \qquad r \geq 1.
\end{equation}
Expressed in terms of the effective Majorana operators, it is given by
\begin{equation}\label{eq:higherisingCharges}
    I_{r+1} = \sum \limits_jh^{(r)}_{j}, \quad h^{(r)}_{j}={\rm i}a_j(\hat{\mathcal{C}}_{j+1}\cdots \hat{\mathcal{C}}_{j+r-1})b_{j+r}.
\end{equation}
Therefore, each local density term spanning sites $j$ through to $j+r$, is bilinear in the Majorana operators $a,b$ supported on the ends, with a string of $\hat{\mathcal{C}}$ operators in between them. We at once identify the above charges with those introduced in \eqref{eq:I_r ising}. For a particular $r$, the local density term satisfies the algebra
\begin{equation}
    h^{(r)}_j h^{(r)}_{{j'}}  = (-1)^{\delta_{|j-{j'}|,r}}
    h^{(r)}_{{j'}} h^{(r)}_j.
\end{equation}
This indicates that the local densities commute at every separations except
for relative distance $r$, where they anti-commute. We note that, this is precisely the algebra we obtained in \eqref{eq:shifted-Ising-algebra}. 

\subsection{Spin-1/2 conserved quantities from fermionic charges}
\label{sec:ferm-spin}
So far we have discussed the integrable structure of the antiperiodic fermionic Hamiltonian $H_-$ \eqref{eq:PBC_APBC} and systematically extracted the tower of conserved charges.
We now demonstrate how these fermionic charges help construct the appropriate integrals of motion for the local spin-1/2 Hamiltonian \eqref{eq:(6,3)-H} 
\begin{eqnarray}
H=\sum_{j=1}^Nh^z_j+\sum_{j=1}^{N}h^{xx}_j.
\end{eqnarray}
On an infinite chain, one can use the JW transformation \eqref{eq:JW1-transformation} and the global rotation \eqref{eq:rotation} to map either of $H_\pm$ \eqref{eq:PBC_APBC} to the above Hamiltonian $H$. Subsequently, all the conserved charges derived for $H_-$ can be mapped to well-defined spin-1/2 integrals of motion. However, this is not quite the case for a finite lattice. As one can verify directly, although the fermionic Hamiltonians $H_\pm$ in \eqref{eq:PBC_APBC} reproduce the Hamiltonian $H$ in the bulk, neither agrees with the spin Hamiltonian at the boundary. In particular, both $H_\pm$ acquire non-local boundary terms when represented in terms of spin-1/2 variables as

\begin{eqnarray}
RH_\pm R^{-1}=\sum_{j=1}^Nh^z_j+\sum_{j=1}^{N-1}h^{xx}_j\mp\hat{\cal P}h^{xx}_N.
\end{eqnarray}
Notably, we rotated the fermionic Hamiltonians by the operator $R$ according to the relation \eqref{eq:rotation}. In other words, the JW transform introduces nonlocal boundary terms in both $H_\pm$, showing that neither of them coincides exactly with the spin-1/2 Hamiltonian $H$ over the complete Hilbert space. Interestingly, since both $H_\pm$ preserve parity, we can still express the local spin Hamiltonian as 
\begin{widetext}
\begin{eqnarray}
    H=\hat{\cal P}_+\left(RH_-R^{-1}\right)\hat{\cal P}_++\hat{\cal P}_-\left(RH_+R^{-1}\right)\hat{\cal P}_-,\qquad \hat{\cal P}_\pm=\frac{1\pm\hat{\cal P}}{2},
\end{eqnarray}
Here $\hat{\cal P}$ is the parity operator defined in \eqref{eq:spin-parity}. It follows immediately that, given the conserved quantities $Q_\pm$ for the fermionic Hamiltonians $H_\pm$, we can construct the conserved charges for the Hamiltonian $H$ as
\begin{eqnarray}\label{eq:QfromQ-Q+}
    Q=\hat{\cal P}_+\left(RQ_-R^{-1}\right)\hat{\cal P}_++\hat{\cal P}_-\left(RQ_+R^{-1}\right)\hat{\cal P}_-,\qquad[Q_\pm,H_\pm]=0,\qquad [Q,H]=0.
\end{eqnarray}
Note that, for charges $Q_\pm$ which anticommute with the parity $\hat{\cal P}`$ and hence do not act within a fixed parity subspace, the corresponding contribution to the spin-1/2 charges vanishes.

However, since the integrable construction described above yields $H_-$ only, it is natural to ask whether a conserved charge $Q$ for the spin-1/2 Hamiltonian $H$ can be constructed solely from the corresponding $Q_-$. To this end, recall that $H_\pm$ are related to each other by a local conjugation by $\gamma^\varepsilon$ \eqref{eq:PBC_APBC_Equiv}, leading to 
\begin{eqnarray}\label{eq:HfromH-H+}
    H=\hat{\cal P}_+\left(RH_-R^{-1}\right)\hat{\cal P}_++\hat{\cal P}_-\left(R\gamma^\varepsilon\right)H_-\left(R\gamma^\varepsilon\right)^{-1}\hat{\cal P}_-,\qquad \sum_{k=1}^3\varepsilon_k\omega_k=0.
\end{eqnarray}
From the above relation, it becomes clear that both the spin and the fermionic Hamiltonians share the same spectrum. More importantly, given a charge $Q_-$, one can now construct 
\begin{eqnarray}\label{eq:Q+fromQ-}
    Q_+=\left(\gamma^\varepsilon\right)Q_-\left(\gamma^\varepsilon\right)^{-1},\qquad [Q_+,H_+]=0,
\end{eqnarray}
which commutes with $H_+$ and hence is a constant of motion for the periodic fermionic Hamiltonian $H_+$. Plugging the expression of the $Q_+$ from \eqref{eq:Q+fromQ-} into the general formula \eqref{eq:QfromQ-Q+} yields
\begin{eqnarray}\label{eq:QfromQ-}
    Q=\hat{\cal P}_+\left(RQ_-R^{-1}\right)\hat{\cal P}_++\hat{\cal P}_-\left(R\gamma^\varepsilon\right)Q_-\left(R\gamma^\varepsilon\right)^{-1}\hat{\cal P}_-,\qquad[Q,H]=0.
\end{eqnarray}
It is straightforward to check that the above charge $Q$ indeed commutes with the Hamiltonian $H$ in \eqref{eq:HfromH-H+} and hence is a valid integral of motion of the spin-1/2 Hamiltonian
\eqref{eq:(6,3)-H}. Among the conserved charges, a particularly important one is obtained from the transfer matrix $\tau(0)$ as
\begin{eqnarray}
    {\cal T}=\hat{\cal P}_+\left(R\tau(0)R^{-1}\right)\hat{\cal P}_++\hat{\cal P}_-\left(R\gamma^\varepsilon\right)\tau(0)\left(R\gamma^\varepsilon\right)^{-1}\hat{\cal P}_-,\qquad {\cal T}^\dagger{\cal T}=1.
\end{eqnarray}
A straightforward calculation shows that ${\cal T}$ acts on the local densities \eqref{eq:hzx} as
\begin{eqnarray}
    {\cal T}h_j^z{\cal T}^{-1}=h_j^{xx},\qquad {\cal T}h_j^{xx}{\cal T}^{-1}=h_{j+1}^{z}. 
\end{eqnarray}
Consequently, for arbitrary coupling $g$ in \eqref{eq:(6,3)-H}, we have
\begin{eqnarray}
    {\cal T}H_g{\cal T}^{-1}=gH_{1/g}.
\end{eqnarray}
Thus ${\cal T}$ realizes an invertible \textit{Kramers-Wannier}-like duality \cite{KW1941-1,KW1941-2}. As a result, the regions $g<1$ and $g>1$ possess identical spectral structure. This is in sharp contrast to the conventional Ising chain, where the corresponding duality is intrinsically non-invertible \cite{aasen2016topological,aasen2020topological,seiberg2024majorana,Sinha:2025wqf,Sinha:2025tqg} and the spectral properties on the both sides of the self-dual point are completely different.
\end{widetext}

\section{Hamiltonians with finite range interactions}
\label{sec:Long-range Hamiltonians}
Interestingly, the fermionic operators $a_j,b_j$ allow us to write an infinite number of non-interacting theories, with the general structure
\begin{eqnarray}
    H={\rm i}\sum_{j,k}f_{jk}(\hat{\cal C}){b}_{j}{b}_k,
\end{eqnarray}
where $f_{jk}(\hat{\cal C})$ is an arbitrary function of the central elements $\{\hat{\cal C}_k\}$. In particular, we shall be interested in the translation invariant, long-range Hamiltonians
\begin{eqnarray}
    H^{(r)}=\sum_{j=1}^{2N} h_j^{(r)},\qquad h_{j}^{(r)}={\rm i}a_jb_{j+r}={\rm i}\hat{\cal C}_jb_jb_{j+r},
\end{eqnarray}
which amounts to consider $f_{jk}^{(r)}(\hat{\cal C})=\delta_{k-j,r}\hat{\cal C}_j$. The local densities now satisfy the algebra
\begin{eqnarray}\label{eq:shifted-Ising-algebra}
    h^{(r)}_jh^{(r)}_k=(-1)^{\delta_{|j-k|,r}}h^{(r)}_kh^{(r)}_j.
\end{eqnarray}
Clearly, for $r=1$ we recover our nearest neighbor Majorana chain. However, for a generic $r$, the action of $H^{(r)}$ partitions the $2N$ indices into several disjoint sublattices. To see this, let us define
\begin{eqnarray}
    {\cal G}:={\rm GCD}(2N,r).
\end{eqnarray}
Here GCD denotes the greatest common divisor. 
Then it is easy to check that, under the shift by $r$-sites, the lattice breaks up into the disconnected sublattices
\begin{eqnarray}
    {\cal S}_p=\left\{p+\mu r\,\Big| \,\mu=0,1,\cdots,\frac{2N}{\cal G}-1\right\},
\end{eqnarray}
with $p=1,\cdots,{\cal G}$. In other words, an index $j_p$ in the sublattice ${\cal S}_p$ satisfies
\begin{eqnarray}
    j_p=p~{\rm mod}({\cal G})\in{\cal S}_p.
\end{eqnarray}
This establishes that different sublattices do not share any common index and hence are mutually disjoint. Evidently, for ${\cal G}=1$, i.e. when $2N$ and $r$ are coprimes, there is no sublattice structure and we always can relabel the Majorana indices so that it becomes like the nearest neighbour Hamiltonian \eqref{eq:H_c} and therefore has exactly the same spectrum as that of \eqref{eq:H_c}. When ${\cal G}\neq 1$, the total Hamiltonian is given by
\begin{eqnarray}\label{eq:long-range-decom}
    &&H^{(r)}=\sum_{p=1}^{\cal G}H^{(r|p)},\nonumber\\
    H^{(r|p)}&=&{\rm i}\sum_{\mu=0}^{\frac{2N}{\cal G}-1}\hat{\cal C}_{p+\mu r}b_{p+\mu r}b_{p+(\mu+1)r},
\end{eqnarray}
where each $H^{(r|p)}$ acts as a nearest neighbor Majorana Hamiltonian, acting non-trivially on the sublattice ${\cal S}_p$. Then it is easy to find the spectrum and the corresponding degeneracies. Notice that, since there is no overlap among the sublattices, we have
\begin{eqnarray}
     \left[H^{(r|p)},H^{(r|p')}\right]=0,
\end{eqnarray}
and the complete Hilbert space splits as
\begin{eqnarray}
   &{\cal H}=\bigotimes_{p=1}^{\cal G}{\cal H}_{p},\qquad {\rm dim}\,{\cal H}_{p}=2^{3N/{\cal G}},
\end{eqnarray}
with ${\cal H}_{p}$ furnishing a representation of the fermions $a_j,b_j,\,j\in{\cal S}_p$. We implicitly assumed that $2N/{\cal G}$ is even. The individual ${\cal H}_p$ further breaks into direct sums according to the eigenvlues of $\{\hat{\cal C}_{p+\mu r}\}$ as
\begin{eqnarray}
    {\cal H}_p=\bigoplus_{\{c_{p+\mu r}\}}{\cal H}_{\{c_{p+\mu r}\}},\qquad {\rm dim}\,{\cal H}_{c_{p+\mu r}}=2^{N/{\cal G}},
\end{eqnarray}
with the Hamiltonian
\begin{eqnarray}
    H_{\{c_{p+\mu r}\}}^{(r|p)}={\rm i}\sum_{\mu=0}^{\frac{2N}{\cal G}-1}c_{p+\mu r}b_{p+\mu r}b_{p+(\mu+1)r},
\end{eqnarray}
acting on the sector ${\cal H}_{\{c_{p+\mu r}\}}$. 
Following the approach outlined in \eqref{sec:spectrum}, one now can show that the Hamiltonian $H_{\{c_{p+\mu r}\}}^{(r|p)}$ has exactly two inequivalent spectra
\begin{eqnarray}
    {\cal E}^{(r|p)}_\pm({n}^{\cal G}_\pm)=4\sin\left(\frac{{\cal G}\pi {n}^{\cal G}_\pm}{N}\right),
\end{eqnarray}
with the momentum numbers
\begin{eqnarray}
    {n}^{\cal G}_+=0,1,\cdots,\frac{N}{\cal G}-1,\frac{N}{\cal G},~{n}^{\cal G}_-=\frac{1}{2},\cdots,\frac{N}{\cal G}-\frac{1}{2}.
\end{eqnarray}
The total energies and the corresponding degeneracies of the Hamiltonian $H^{(r)}$ then can be obtained from \eqref{eq:long-range-decom}. We do not wish to pursue this line of investigation any further. 
We conclude by noting that the higher Ising charges can also be obtained by considering
\begin{eqnarray}\label{eq:I_r ising}
    f_{jk}^{(r)}(\hat{\cal C})&=&\delta_{k-j,r}~\hat{\cal C}_{j}\cdots\hat{\cal C}_{k-1}.
\end{eqnarray}
One then finds that the above choice yields the mutually commuting conserved quantities in \eqref{eq:higherisingCharges}.


\section{A prescription to generate hidden TFIM's}
\label{sec:gYBE-HFF}
The derivation of the multi-site Hamiltonian in \eqref{eq:(6,3)-H} from the $(d,6,3)$-gYBE suggests that similar multi-site Hamiltonians can be obtained by systematically solving the $(d,2k,k)$-gYBE with Majorana $R$-matrices, such as the one in \eqref{eq:(6,3)-R-1+3}. These generalized $R$-matrices will result in spin chains with multi-site interactions on a total of $2kN$ sites. Their Hamiltonian densities will obey the algebra in \eqref{eq:Ising_algebra} and thus they can be solved using methods outlined in Section \ref{sec:spectrum}. We will now see how to systematically construct such solutions. This amounts to finding representations of the algebra of the Hamiltonian densities in \eqref{eq:Ising_algebra}.

For an arbitrary $k$, we look for Majorana $R$-matrices that solve the $(d,2k,k)$-gYBE with ans\"{a}tze similar to the one in \eqref{eq:(6,3)-R-1+3}. We will use the following notation for the operators $\mathcal{M}$:
\begin{eqnarray}\label{eq:notation-M}
  \mathcal{M}_j \equiv\mathcal{M}_{(jk+1)\cdots (jk+2k)},
\end{eqnarray}
with $j\in\{0,1,\cdots,2(N-1)\}$. As required by the ans\"{a}tze for the generalized $R$-matrices, the $\mathcal{M}$'s have support on $2k$ indices. Consecutive $\mathcal{M}$'s are shifted by units of $k$, so that the former can possibly satisfy the $(d,2k,k)$-gYBE. The conditions on $\mathcal{M}$ for the latter are precisely,
\begin{eqnarray}
   & \{\mathcal{M}_j, \mathcal{M}_m\} =2\delta_{jm},\quad|j-m|=1 & \nonumber \\
   & \left[\mathcal{M}_j, \mathcal{M}_m\right] = 0,\quad |j-m|>1.&
\end{eqnarray}
We are interested in Majorana representations of this algebra such that the resulting Hamiltonian is quadratic in the effective Majoranas, \eqref{eq:H-ff-form}. This ensures that the equivalent multi-site spin Hamiltonians are hidden TFIM's. This is achieved by choosing the $\mathcal{M}$ as:
\begin{eqnarray}
    \mathcal{M}_j = \mathrm{i}\,a_j b_{j+1},
\end{eqnarray}
with $a$ and $b$ being effective Majoranas. The supports of $a$ and $b$ together span $2k$ indices of the Majoranas, $\gamma$ [See \eqref{eq:notation-M}]. In some cases they have equal support of $k$, $\gamma$ Majorana indices, and in some cases they have unequal support as we shall soon see. This choice for $\mathcal{M}$ takes us from the original $2kN$ Majorana indices to $2N$ effective Majorana indices. With this choice, the algebra of $\mathcal{M}$'s in \eqref{eq:M-algebra} is then guaranteed to be satisfied as these effective Majoranas, $a$ and $b$ satisfy the relations in \eqref{eq:effective-Majorana-relations}. For instance, the anticommutativity follows from a short computation, 
\begin{eqnarray}
    \mathcal{M}_j\mathcal{M}_{j+1}=-a_ja_{j+1}b_{j+1}b_{j+2}=-\mathcal{M}_{j+1}\mathcal{M}_j.
\end{eqnarray}
Similar computations using the relations in \eqref{eq:effective-Majorana-relations} prove the other identities of \eqref{eq:M-algebra}. These arguments show that representations of the algebra \eqref{eq:M-algebra} reduces to finding a realization of the effective Majoranas, $a$ and $b$, in terms of the $2k$ Majoranas, $\gamma$. 

The procedure is as follows. First, we split the $2k$ indices on $\mathcal{M}$ into two blocks, each having $k$ consecutive indices. Each of these blocks will be further split into smaller parts such that the total number of parts is even. These criteria have to be met for the QISM to go through for the generalized $R$-matrices [See Section \ref{sec:integrability}]. A few examples will illustrate this procedure in a satisfactory manner. We will consider the odd and even $k$'s separately, beginning with the odd values of $k$. 

\subsection{Odd $k$}
\label{subsec:odd-k}
The simplest value is $k=1$. This corresponds to 
\begin{eqnarray}
    \mathcal{M}_{12} = \mathrm{i}\,\gamma_1\gamma_2.
\end{eqnarray}
This results in a $R$-matrix that solves the $(d,2,1)$-gYBE, which is just the usual YBE. In the spin representation this yields precisely the TFIM, \eqref{eq:Ham_TFIM} at criticality. For details see \cite{Sinha:2025wqf}.

The next non-trivial value is when $k=3$. In this case the 6 Majoranas $\gamma$'s can be split into 2, 4 or 6 parts. The different choices are shown in Table \ref{tab:M-k-3}. All the coefficients $\omega$'s are real numbers. In each case we can normalize $\mathcal{M}$ by diving them by the square root of the sum of the squares of the $\omega$'s in order to ensure that $\mathcal{M}^2=\mathbb{1}$. For instance in the $1+1$ case we divide by $\left(\sum_{j=1}^3\omega_j^2 \right)^{1/2}\left(\sum_{j=4}^6\omega_j^2 \right)^{1/2}$.
    \begin{table*}[t]
\caption{Possible Majorana realizations of $\mathcal{M}_{1\cdots6}$ for $k=3$. 
The terminology ``The split'' corresponds to partitions of integers. 
We avoid using the word ``partition'' so as not to confuse it with the statistical-mechanical terminology of ``partition function''.}
\label{tab:M-k-3}

\begin{tabular}{c|c|c}
\hline\hline
\textbf{No. of parts} & \textbf{The split} & $\boldsymbol{\mathcal{M}}$ \\
\hline

2 & $1+1$ &
$\mathrm{i}
\left(\sum\limits_{j=1}^{3}\omega_j\gamma_j\right)
\left(\sum\limits_{j=4}^{6}\omega_j\gamma_j\right)$
\\
\hline

\multirow{3}{*}{4}
& $1+3$ &
$\left(\sum\limits_{j=1}^{3}\omega_j\gamma_j\right)
\gamma_4\gamma_5\gamma_6$
\\
\cline{2-3}

& $3+1$ &
$\gamma_1\gamma_2\gamma_3
\left(\sum\limits_{j=4}^{6}\omega_j\gamma_j\right)$
\\
\cline{2-3}

& \rule{0pt}{16pt} $2+2$ &
$\left(\omega_1\gamma_1+\omega_2\gamma_2\right)
\gamma_3
\left(-\omega_2\gamma_4+\omega_1\gamma_5\right)
\gamma_6$
\\[8pt]
\hline

6 & $3+3$ &
$\mathrm{i}\prod\limits_{j=1}^{6}\gamma_j$
\\
\hline\hline
\end{tabular}
\end{table*}

\begin{widetext}
The Hamiltonian for $1+1$ case is given by 
\begin{eqnarray}
    H^{(1+1)}=\mathrm{i}\sum\limits_{j} \left(\omega_1\gamma_{3j-2}+\omega_2\gamma_{3j-1}+\omega_3\gamma_{3j}\right)\left(\omega_4\gamma_{3j+1}+\omega_5\gamma_{3j+2}+\omega_6\gamma_{3j+3}\right).
\end{eqnarray}
Using the JW transformation \eqref{eq:JW1-transformation}, we can find the spin-1/2 representation of this model as 
\begin{eqnarray}\label{eq:k-6-H-1+1}
    H^{(1+1)}&=&\sum\limits_{j} (\omega_1Y_{3j-2}-\omega_2X_{3j-2})(\omega_4Y_{3j-1}+\omega_5Z_{3j-1}X_{3j}+\omega_6Z_{3j-1}Y_{3j})-\omega_3\omega_4Z_{3j-1}+\omega_3Y_{3j-1}(\omega_5X_{3j}+\omega_6Y_{3j})\nonumber\\
    &+&\sum\limits_{j}(-\omega_1X_{3j-1}Z_{3j}+\omega_2Y_{3j}-\omega_3X_{3j})(\omega_4X_{3j+1}+\omega_5Y_{3j+1}+\omega_6Z_{3j+1}X_{3j+2}).
\end{eqnarray}
The above Hamiltonian is defined strictly over an infinite chain. For a finite lattice, the local fermionic Hamiltonian acquires a non-local spin-1/2 boundary term after the JW transformation. However, this does not affect the bulk Hamiltonian densities. Similarly, for $(3+1)$, $(2+2)$ and $(3+3)$, we have the Hamiltonians
\begin{subequations}
    \begin{eqnarray}\label{eq:k-6-H-3+1}
    H^{(3+1)}&=&\sum\limits_{j}\gamma_{3j-2}\gamma_{3j-1}\gamma_{3j}(\omega_1\gamma_{3j+1}+\omega_2\gamma_{3j+2}+\omega_3\gamma_{3j+3})\nonumber\\
    &=&-\sum\limits_{j} Z_{3j-2}(\omega_1Z_{3j-1}-\omega_2Y_{3j-1}X_{3j}-\omega_3Y_{3j-1}Y_{3j})\nonumber\\&&-\sum\limits_{j}X_{3j-1}(\omega_1X_{3j+1}+\omega_2Y_{3j+1}+\omega_3Z_{3j+1}X_{3j+2}).
    \label{eq:H-(2+2)}
    \\
    H^{(2+2)}&= &\sum\limits_{j} \left(\omega_1\gamma_{3j-2}+\omega_2\gamma_{3j-1}\right)\gamma_{3j}\left(-\omega_2~\gamma_{3j+1}+\omega_1\gamma_{3j+2}\right)\gamma_{3j+3}\nonumber\\
    &=&\sum\limits_{j}\omega_1\omega_2( Y_{3j-2}Y_{3j}+X_{3j-1}Y_{3j}Y_{3j+1}X_{3j+2}+X_{3j-2}X_{3j-1}Z_{3j}+Z_{3j}X_{3j+1}X_{3j+2})\nonumber\\
    &&\qquad-\omega_1^2(Y_{3j-2}X_{3j-1}Z_{3j}-X_{3j-1}Y_{3j}X_{3j+1}X_{3j+2})-\omega_2^2(X_{3j-2}Y_{3j}-Z_{3j}Y_{3j+1}X_{3j+2}).
    \label{eq:k-6-H-3+3}
    \\
    H^{(3+3)}&=&\mathrm{i}\sum\limits_{j} \gamma_{3j-2}\gamma_{3j-1}\gamma_{3j}\gamma_{3j+1}\gamma_{3j+2}\gamma_{3j+3}=-\sum\limits_{j} Z_{3j-2}Z_{3j-1}Z_{3j}-\sum\limits_{j}X_{3j-1}X_{3j+2}.
\end{eqnarray}
\end{subequations}
\end{widetext}
All the above the Hamiltonians corresponding to the different splits can be solved using the effective Majorana technique outlined in Section \ref{sec:spectrum}. In each case the effective Majoranas, $a$ and $b$, written in terms of the original $\gamma$ Majoranas, are chosen such that they   satisfy the relations in \eqref{eq:effective-Majorana-relations}. The difference arises between the {\it odd-odd} split ($1+1$, $3+1$, $3+3$) and the {\it even-even} split ($2+2$). In the odd-odd split the effective Majorana $a$ is mapped to the $\gamma$ Majoranas with support on the first 3 sites, whereas the effective Majorana $b$ is mapped to the $\gamma$ Majoranas with support on the next 3 sites. Thus, from Table \ref{tab:M-k-3}, we can write down the $\mathcal{M}_1=\mathrm{i}\,a_1b_{2}$ or $\mathcal{M}_1=\mathrm{i}\,a_1a_{2}$ for each of these cases as
\begin{eqnarray}\label{eq:1,3:3,1}
    & 1+1 & :~~a_1 = \sum\limits_{j=1}^3\omega_j\gamma_j,\quad a_{2} = \sum\limits_{j=4}^6~\omega_j\gamma_j,  \nonumber \\
    & 1+3 & :~~a_1 = \sum\limits_{j=1}^3\omega_j\gamma_j,\quad b_2 = \gamma_4\gamma_5\gamma_6, \nonumber \\
    & 3+3 & :~~a_1 = \gamma_1\gamma_2\gamma_3,\quad a_2=\gamma_4\gamma_5\gamma_6.   
\end{eqnarray}
On the other hand, in the even-even split the support of the effective Majorana, $a$ will exceed the first 3 sites. Then the effective Majorana, $b$ will be supported on the remaining sites which is less than 3. For instance in the $2+2$ case from Table \ref{tab:M-k-3} we have 
\begin{eqnarray}
    2+2 : a_1 = \left(\omega_1\gamma_1 +\omega_2\gamma_2\right)\gamma_3\left(-\omega_2\gamma_4 +\omega_1\gamma_5\right),~ b_2=\gamma_6.\nonumber\\
\end{eqnarray}
As mentioned earlier, this choice is required for these effective Majoranas to satisfy the relations in \eqref{eq:effective-Majorana-relations}. We expect this difference in choices of effective Majoranas  between the odd-odd and even-even splits to continue for higher $k$ as well.

\begin{figure}[h!]
    \centering
    \includegraphics[width=0.6\linewidth]{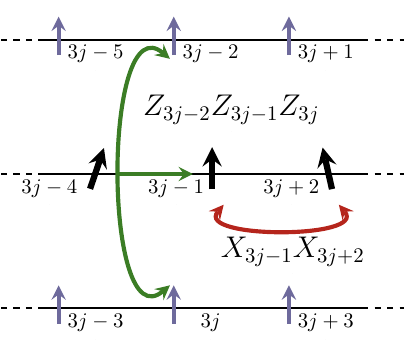}
    \caption{Action of the Hamiltonian density of $H^{(3+3)}$. The grey spins in the first and third layer are non-dynamical and hence act as a classical background field for the dynamical spins governed by the TFIM Hamiltonian in the second layer.}
    \label{fig:H-3+3}
\end{figure}

At this point we should also mention that the spin Hamiltonian in \eqref{eq:k-6-H-3+3} corresponding to the split $3+3$ is particularly simple to solve in the spin basis as well. To see this, we observe that the lattice of $3N$ sites can be split into three layers, with the indices, $j$ of the $l$-th layer satisfying $j~\textrm{mod}~3=l$, $l\in\{1,2,3\}$. Thus the first layer contains the sites $\{1,4,7\cdots\}$, the second consists of $\{ 2,5,8,\cdots\}$ and the third layer contains the multiples of 3, $\{3,6,9,\cdots\}$. The action of the Hamiltonian densities can now be schematically shown, Figure \ref{fig:H-3+3}.
It is then clear from the Hamiltonian in \eqref{eq:k-6-H-3+3}, that the operators $Z_j$, with the indices $j$ taking values in the first and third layers, are conserved charges. This makes every eigenvalue $2^{2N}$-fold degenerate. Thus the spins located in these two layers are non-dynamical under the unitary time evolution generated by \eqref{eq:k-6-H-3+3}. Let us denote the eigenvalue of $Z_{3j-2}Z_{3j}$ to be $g_j$. Then $g_j$ takes values $\pm 1$. The $2^{2N}$ sectors are labeled by the eigenvalues of $g_j$, for $j=1,\cdots, N$. This block diagonalizes the Hamiltonian \eqref{eq:k-6-H-3+3}. Within each block the Hamiltonian is unitarily equivalent to a TFIM. There are two possibilities. When $g_j=1$, the spins on sites $3j-2$ and $3j$ are both either +1 or -1 eigenstates of $Z$. In this case the Hamiltonian density for $j$ coincides with the Hamiltonian density of the TFIM. On the other hand, when $g_j=-1$, then the spins on $3j-2$ and $3j$ point in opposite directions. In this case the Hamiltonian density is unitarily equivalent to the Hamiltonian density of the TFIM, with the unitary mapping given by $X_{3j-1}$. These two cases are sufficient to determine the full structure of the Hamiltonian \eqref{eq:k-6-H-3+3}, as the spins on sites $3j-2$ and $3j$, do not interact with its neighbors on the first and third layers. The ground state is then a direct sum of the TFIM ground states in each sector labeled by the eigenvalues of $g_j$, each of which is obtained with an appropriate action of the Pauli $X_{3j-1}$'s on the ground state of the standard TFIM. An immediate consequence of the block diagonal structure of the Hamiltonian is that the $Z$ correlators in this theory vanish when computed in the ground state of this model. As a special case, this implies that the magnetization in this model is 0 when compared to the standard TFIM. Furthermore, we expect the entanglement entropy of the ground states of this system to also account for the uncertainty in choosing the sectors labeled by eigenvalues of $g_j$. This analysis will be carried in a future work. We end this discussion by noting that the fate of the non-invertible KW duality-symmetry in the present case is expected differ from that in the $1+3$-case. One indication is that the Majorana Hamiltonians ${\rm i}\sum_ja_ja_{j+1}$ and ${\rm i}\sum_j(-1)^{\delta_{j,2N}}a_ja_{j+1}$ for the $3+3$ case have completely different spectra altogether.

The number of choices for $\mathcal{M}$ and the associated multi-site Hamiltonians drastically increase with $k$. We will not attempt to write down the possible $\mathcal{M}$ choices for an arbitrary value of odd $k$. Nevertheless, we will consider the $k=5$ case in Appendix \ref{app:k=5} to give an idea of the type of multi-site interactions that can occur for higher values of odd $k$. This example will illustrate the algorithm used to construct the Majorana representations of the algebra of $\mathcal{M}$ operators in \eqref{eq:M-algebra}.

\subsection{Even $k$}
\label{subsec:even-k}
We now consider even $k$ representations of $\mathcal{M}$ beginning with 4-site $\mathcal{M}$'s or $k=2$. In this case the total number of parts can be either 2 or 4, but only the former leads to a $\mathcal{M}$ that satisfies \eqref{eq:M-algebra}\footnote{For arbitrary even $k$, $\mathcal{M}$ cannot be split into $2k$ parts as the resulting $\mathcal{M}$ does not satisfy the required algebra, \eqref{eq:M-algebra}. Thus this case will not be considered while discussing other even $k$ values. }. This is given by
\begin{eqnarray}
    \mathcal{M}_{1\cdots4} = {\rm i}\left(\omega_1\gamma_1 + \omega_2\gamma_2\right)\left(\omega_3\gamma_3 + \omega_4\gamma_4\right),    
\end{eqnarray}
with the condition $\omega_1\omega_4-\omega_2\omega_3=0$. The Majorana and spin Hamiltonians corresponding to this $\mathcal{M}$ are
\begin{widetext}
\begin{eqnarray}
    H &=&\mathrm{i}\sum\limits_{j}(\omega_1\gamma_{2j-1}+\omega_2\gamma_{2j})(\omega_3\gamma_{2j+1}+\omega_4\gamma_{2j+2})=-\sum\limits_{j} (\omega_1Y_j-\omega_2X_j)(\omega_3X_{j+1}+\omega_4Y_{j+1}).
\end{eqnarray}
This model can be rotated to the familiar looking spin Hamiltonian $H = \sum_j\tilde{X}_j\tilde{Y}_{j+1}$.

There are more possibilities for $k=4$. In this case $\mathcal{M}$ can be split into either 2,4, or 6 parts. The only split for the 2 part case, is $1+1$. The $\mathcal{M}$ operator can then be realized in two ways,
\begin{eqnarray}
    \mathcal{M}_{1\cdots 8} = \begin{cases}
\text{Case I}:\quad\left(\sum\limits_{j=1}^4~\omega_j\gamma_j\right)\left(\sum\limits_{j=1}^4~\omega_j\gamma_{j+4}\right), \\
\text{Case II}:\quad\left(\omega_1\gamma_1\gamma_2\gamma_3 + \omega_2\gamma_4\right)\left(\omega_3\gamma_5\gamma_6\gamma_7 + \omega_4\gamma_8\right),\quad\omega_1\omega_4 - \omega_2\omega_3=0.
    \end{cases}
\end{eqnarray}
The explicit forms of the two Majorana Hamiltonians are given, respectively, by
\begin{eqnarray}
    H^{\rm I} &=&\mathrm{i}\sum\limits_{j}(\omega_1\gamma_{4j-3}+\omega_2\gamma_{4j-2}+\omega_3\gamma_{4j-1}+\omega_4\gamma_{4j})(\omega_5\gamma_{4j+1}+\omega_6\gamma_{4j+2}+\omega_7\gamma_{4j+3}+\omega_6\gamma_{4j+4}),\nonumber\\ 
    &=&\sum_{j}(\omega_1Y_{2j-1}Z_{2j}-\omega_2 X_{2j-1}Z_{2j}+\omega_3 Y_{2j}-\omega_4 X_{2j})(\omega_5 X_{2j+1}+\omega_6 Y_{2j+1}+\omega_7 Z_{2j+1}X_{2j+2}+\omega_8 Z_{2j+1}Y_{2j+2}),\nonumber\\
    \\
    H^{\rm II}&=&\sum\limits_{j}\left(\omega_1\gamma_{4j-3}\gamma_{4j-2}\gamma_{4j-1} + \omega_2\gamma_{4j}\right)\left(\omega_3\gamma_{4j+1}\gamma_{4j+2}\gamma_{4j+3} + \omega_4\gamma_{4j+4}\right)\nonumber\\
    &=&\sum_{j}(\omega_1 Z_{2j-1}Y_{2j}+\mathrm{i}\omega_2 X_{2j})(\mathrm{i}\omega_3 X_{2j+2}+\omega_4 Z_{2j+1}Y_{2j+2}).
\end{eqnarray}
Moving to the 4 part case, we consider the splits $3+1$ or $1+3$ and $2+2$. For the $3+1$ split we have two possible $\mathcal{M}$ operators given by 
\begin{eqnarray}
    \mathcal{M}_{1\cdots 8} = \begin{cases}\text{Case III}:\quad
\left(\omega_1\gamma_1+\omega_2\gamma_2\right)\gamma_3\gamma_4\left(\sum\limits_{j=3}^6\omega_j\gamma_{j+2}\right),\quad\omega_1\omega_4-\omega_2\omega_3=0, \\
\text{Case IV}:\quad\left(\omega_1\gamma_1 + \omega_2\gamma_2\right)\gamma_3\gamma_4\gamma_8. 
    \end{cases}
\end{eqnarray}
The Hamiltonians associated with these $\mathcal{M}$'s are given by 
\begin{eqnarray}
    H^{\rm III}&=&\sum\limits_{j}  \left(\omega_1\gamma_{4j-3}+\omega_2\gamma_{4j-2}\right)\gamma_{4j-1}\gamma_{4j}(\omega_3\gamma_{4j+1}+\omega_4\gamma_{4j+2}+\omega_5\gamma_{4j+3}+\omega_6\gamma_{4j+4})\nonumber\\
    &=&\sum\limits_{j}(\omega_1Y_{2j-1}-\omega_2X_{2j-1})(\omega_3X_{2j+1}+\omega_4Y_{2j+1}+Z_{2j+1}(\omega_5X_{2j+2}+\omega_6Y_{2j+2}))\\
    H^{\rm IV}&=&\sum\limits_{j}\left(\omega_1\gamma_{4j-3} + \omega_2\gamma_{4j-2}\right)\gamma_{4j-1}\gamma_{4j}\gamma_{4j+4}=\sum\limits_{j} (\omega_1Y_{2j-1}-\omega_2X_{2j-1})Z_{2j+1}Y_{2j+2}.
\end{eqnarray}
The only possible $\mathcal{M}$ for the $2+2$ split is 
\begin{eqnarray}
    \mathcal{M}_{1\cdots 8}(\text{Case V})= \left(\omega_1\gamma_1+\omega_2\gamma_2\right)\left(\omega_3\gamma_3+\omega_4\gamma_4\right)\left(\omega_5\gamma_5+\omega_6\gamma_6\right)\left(\omega_7\gamma_7+\omega_8\gamma_8\right),
\end{eqnarray}
the real coefficients $\omega_j$'s satisfy 
\begin{eqnarray}
(\omega_1\omega_3\omega_5\omega_7+\omega_1\omega_4\omega_5\omega_8+\omega_2\omega_3\omega_6\omega_7+\omega_2\omega_4\omega_6\omega_8)
=0
=(\omega_1\omega_4\omega_6\omega_7 + \omega_2\omega_3\omega_5\omega_8 - \omega_2\omega_4\omega_5\omega_7 - \omega_1\omega_3\omega_6\omega_8).
\end{eqnarray}
This multi-site operator yields a periodic Majorana Hamiltonian, given by 
\begin{eqnarray}
    H^{\rm V}&=&\sum\limits_{j}\left(\omega_1\gamma_{4j-3}+\omega_2\gamma_{4j-2}\right)\left(\omega_3\gamma_{4j-1}+\omega_4\gamma_{4j}\right)\left(\omega_5\gamma_{4j+1}+\omega_6\gamma_{4j+2}\right)\left(\omega_7\gamma_{4j+3}+\omega_8\gamma_{4j+4}\right)\nonumber\\
    &=&\sum\limits_{j} (\omega_1Y_{2j-1}-\omega_2X_{2j-1})(\omega_3X_{2j}+\omega_4Y_{2j})(\omega_5X_{2j+1}-\omega_6Y_{2j+1})(\omega_7X_{2j+2}+\omega_8Y_{2j+2}).
\end{eqnarray}
When $\mathcal{M}$ is broken into 6 parts, the possible splits are $4+2$ or $2+4$ and $3+3$. The first two splits do not result in realizations of the algebra in \eqref{eq:M-algebra}. Thus, we are left with only the $3+3$ split, whose corresponding $\mathcal{M}$ operator is 
\begin{eqnarray}
    \mathcal{M}_{1\cdots 8}(\text{Case VI}) = \left(\omega_1\gamma_1 + \omega_2\gamma_2\right)\gamma_3\gamma_4\left(\omega_3\gamma_5+\omega_4\gamma_6\right)\gamma_7\gamma_8,\quad\omega_1\omega_4 - \omega_2\omega_3=0.
\end{eqnarray}
This multi-site operator corresponds to a anti-periodic Majorana Hamiltonian 
\begin{eqnarray}
    H^{\rm VI}&=&{\rm i}\sum\limits_{j} \left(\omega_1\gamma_{4j-3} + \omega_2\gamma_{4j-2}\right)\gamma_{4j-1}\gamma_{4j}\left(\omega_3\gamma_{4j+1}+\omega_4\gamma_{4j+2}\right)\gamma_{4j+3}\gamma_{4j+4}\nonumber\\
    &=&\sum\limits_{j} (\omega_1Y_{2j-1}-\omega_2X_{2j-1})(\omega_3 X_{2j+1}+\omega_4Y_{2j+1})Z_{2j+2}
\end{eqnarray}
For finding the solution to the Hamiltonian, we could proceed in the same way as the odd-$k$ case in \ref{subsec:odd-k}, where the Hamiltonian can be effectively rewritten in terms of the effective Majorana operators $a$ and $b$. In the odd-odd splitting, the effective Majorana operator $a$ is composed of $k$ consecutive $\gamma$ Majorana operators acting on the first $k$ sites, while the effective Majorana operator $b$ is similarly constructed from the $\gamma$ Majoranas supported on the next $k$ sites. This structure of the Majorana Hamiltonian leads to anti-periodic boundary conditions. One such example is given below 
\begin{eqnarray}
    1+1: \quad a_1 =\sum\limits_{j=1}^4 \omega_j \gamma_{j},\quad b_2 =\sum\limits_{j=1}^4 \omega_j \gamma_{j+4}.
\end{eqnarray}
On the contrary, for the even-even splitting, an unequal number of $\gamma$ Majorana operators need to be used to ensure that the algebra involving the effective Majorana operators $a$ and $b$ holds. The Hamiltonian obtained in this case satisfies periodic boundary conditions. For example,
\begin{eqnarray}
    2+2: \quad a_1 =\left(\omega_1\gamma_1+\omega_2\gamma_2\right)\left(\omega_3\gamma_3+\omega_4\gamma_4\right)\left(\omega_5\gamma_5+\omega_6\gamma_6\right),\quad b_2 =\left(\omega_7\gamma_7+\omega_8\gamma_8\right).
\end{eqnarray}

\end{widetext}

\section{Conclusion}
\label{sec:outlook}
In this work we have introduced one dimensional spin chains with multi-site interactions, that look unsolvable in the spin basis, but are nevertheless free-fermionic after a sequence of transformations, $\Phi_{\rm Spin}=\Phi_{\rm JW_2}\circ\Phi_{\rm Majorana}\circ\Phi_{\rm JW_1}$ as elaborated in Section \ref{subsec:hiddenIsing}. The local Hamiltonian densities obey the Ising exchange-algebra, \eqref{eq:Ising_algebra}. We work with certain representations of this algebra, those which can be decomposed as a product of the effective Majoranas, $\mathcal{M}_j=\mathrm{i}~a_jb_{j+1}$, satisfying a modified CAR algebra, \eqref{eq:effective-Majorana-relations}\footnote{Another mapping of such systems to free-fermions can be found in \cite{minami2016solvable}.}. In particular, these representations have centers [set of elements commuting with everything else in the algebra], generated by local conserved charges $\mathcal{C}_j$ as introduced in \eqref{eq:effective-Majorana-relations}-\eqref{eq:C-properties}. 
Using this, the spectrum of the Hamiltonian, \eqref{eq:(6,3)-H}-\eqref{eq:hzx}, was studied in detail in Section \ref{sec:spectrum} and was seen to coincide with that of the TFIM modulo a huge degeneracy. This degeneracy was attributed to the presence of the local conserved charges, $\mathcal{C}_j$'s. The thermodynamics of the hidden TFIM was almost similar to the standard TFIM with most of the differences originating in the exponential degeneracy of the spectrum. This is elucidated in Section \ref{sec:thermodynamics}.

All of these models can be unified under the framework of the quantum inverse scattering method in Section \ref{sec:integrability} [See Appendix \ref{app:derivation-Hamiltonian} for derivations]. We show that the critical versions [$g=1$ in \eqref{eq:(6,3)-H}] of these hidden TFIM models are all Yang-Baxter integrable, with the difference that the relevant $R$-matrices have support on more than the usual 2 indices and satisfy the so-called generalized Yang-Baxter equation [See Appendix \ref{app:gYBE}], which is a multi-site generalization of the YBE. The associated conserved quantities can then be derived using a multi-site generalization of the boost operator method as shown in Section \ref{sec:integrability}, with derivations in Appendix \ref{app:Boost-derivation}. And finally, the QISM framework helps us to systematically generalize these hidden TFIM's with local Hamiltonian densities supported on a arbitrary number of sites as discussed in Section \ref{sec:gYBE-HFF} [See also Appendix \ref{app:k=5}]. Alternatively, this can be viewed as an elaborate construction of representations of the Ising exchange algebra, \eqref{eq:Ising_algebra}. 

The effective Majoranas can also be used to construct Hamiltonians with finite range interactions as shown in Section \ref{sec:Long-range Hamiltonians}. The Hamiltonian densities now satisfy a shifted version of the Ising exchange algebra \eqref{eq:shifted-Ising-algebra}. We consider two classes of such Hamiltonians. For one of them, the Hamiltonian densities belong to the higher conserved charges of the hidden TFIM and are thus also free-fermionic. The other set of Hamiltonians have no immediate origin from the QISM formalism. Nevertheless, we still show them to be free-fermionic.

From the perspective of this work, we see that the physical models obeying the Ising exchange algebra are associated with the Baxterized version of the extraspecial 2-group solutions of the generalized Yang-Baxter equation \cite{rowell2010extraspecialtwogroupsgeneralizedyangbaxter}. The latter include the 6-site $R$-matrix in \eqref{eq:(6,3)-R-1+3} and its generalizations constructed using the $\mathcal{M}$'s in Section \ref{sec:gYBE-HFF}. A natural question arises now :

{\it Can we generate the local algebra of Fendley's free-fermion in disguise systems as solutions of the generalized Yang-Baxter equation?  }

Our preliminary analysis answers this question in the affirmative. We illustrate this through a simple example. Consider the following set of  operators,
\begin{eqnarray}
    \tilde{\mathcal{M}}_j = \mathrm{i}\left(\omega_1\gamma_j + \omega_2\gamma_{j+1}\right)\gamma_{j+2}\gamma_{j+3}.
\end{eqnarray}
They satisfy the algebra
\begin{eqnarray}
    &\tilde{\mathcal{M}}^2=\mathbb{1},\quad\{\tilde{\mathcal{M}}_j, \tilde{\mathcal{M}}_k\} =0,~~j\neq k\pm 1,\nonumber\\
    &\left[\tilde{\mathcal{M}}_j, \tilde{\mathcal{M}}_{j+1}\right] =0. 
\end{eqnarray}
Such operators generate a solution of the $(d,4,2)$-gYBE with an ansatz similar to the one in \eqref{eq:(6,3)-R-1+3}. Then we find that products of such operators,
\begin{eqnarray}
    h_j &=& \tilde{\mathcal{M}}_j\tilde{\mathcal{M}}_{j+1}\nonumber\\
    &=&- \left(\omega_1\gamma_j + \omega_2\gamma_{j+1}\right)\left(\omega_2\gamma_{j+2} - \omega_1\gamma_{j+3}\right)\times\nonumber\\
    &&\hspace{4.2cm}\gamma_{j+4}\gamma_{j+5},
\end{eqnarray}
satisfy Fendley's free-fermion in disguise algebra \cite{fendley2019free}:
\begin{eqnarray}\label{eq:FFD-algebra}
    &\{h_j,h_k\} = 0,~k=j+1,j+2,\nonumber\\
    &\left[h_j,h_k\right]=0,~k\neq j+1,j+2.
\end{eqnarray}
Furthermore, the operators $\tilde{M}_j$'s are used to construct solutions of the $(d,6,2)$-generalized Yang-Baxter equation whose form is similar to \eqref{eq:(6,3)-R-1+3} [See Appendix \ref{app:gYBE} as well]. We will pursue this direction in a future work.


\begin{acknowledgments}
VK is funded by the U.S. Department of Energy, Office of Science, National Quantum Information Science Research Centers, Co-Design Center for Quantum Advantage ($C^2QA$) under Contract No. DE-SC0012704.
\end{acknowledgments}

\bibliography{refs}
\bibliographystyle{ieeetr}

\onecolumngrid

\appendix

\section{Review of the $(d,l,m)$-gYBE}
\label{app:gYBE}
We denote the generalized $R$-matrix with $l$ indices as,
\begin{eqnarray}\label{eq:genR-matrix}
    R_{1\cdots l}(\left\{u\right\}) \equiv  R_{j_1\cdots j_l}(\left\{u\right\})~;~\left\{u\right\} = \left(u_1, \cdots, u_l\right)\in\mathbb{C}.
\end{eqnarray}
In this notation, the $(d,l,m)$-gYBE reads,
\begin{eqnarray}\label{eq:sp-gYBE}
    R_{1\cdots l}(\left\{u\right\})R_{(1+m)\cdots (l+m)}(\left\{v\right\})R_{1\cdots l}(\left\{w\right\}) = R_{(1+m)\cdots (l+m)}(\left\{w\right\}) R_{1\cdots l}(\left\{v\right\})R_{(1+m)\cdots (l+m)}(\left\{u\right\}). 
\end{eqnarray}
This is an operator equation on $\bigotimes\limits_{j=1}^{l+m}\mathcal{H}_j^d$, with $\mathcal{H}$ being the local Hilbert space. The positive integers, $d$, $l$ and $m$ denote the dimension of the local Hilbert space, the number of indices on the generalized $R$-matrix and the spacing between the indices of two consecutive generalized $R$-matrices in the gYBE, respectively. For a given $l$, we have $m\in\{1,\cdots,l-1\}$. It is clear that for $m\geq l$, the $R$-matrices in the gYBE commute with each other\footnote{Nevertheless, they can be solved and have interesting applications in quantum entanglement theory \cite{Padmanabhan2019QuantumES}.}. The usual YBE is recovered when we set $l=2$ and $m=1$. Note that the index structure of the gYBE is similar to the braid relation 
\begin{eqnarray}\label{eq:braid-relations}
    \sigma_i\sigma_{i+1}\sigma_i = \sigma_{i+1}\sigma_i\sigma_{i+1}~;~\sigma_i\equiv\sigma_{i,i+1}.
\end{eqnarray}
This implies that the constant form of the generalized $R$-matrix, obtained when the spectral parameter is dropped, furnishes a representation of some $N$-strand braid group\footnote{Constant gYBE solutions play a role as entangling quantum gates \cite{Padmanabhan_2020,padmanabhan2020generating,kauffman2018topologicalaspectsquantumentanglement,Kauffman_2004,chen2011generalizedyangbaxterequationsbraiding,Alagic_2016,vasquez2016qubitrepresentationsbraidgroups}}.

We will now consider a class of solutions of the gYBE that is the source of the 6-site $R$-matrix in \eqref{eq:(6,3)-R-1+3}. Consider the $2k$-site generalized $R$-matrices,
\begin{equation}\label{eq:R-gYBE}
    R_{1\cdots 2k}(\lambda) = \mathbb{1} + a(\lambda)~\mathcal{M}_{1\cdots 2k}\;,
\end{equation}
which is composed of a multi-site Majorana operator $\mathcal{M}$ acting on $2k$ number of local Hilbert spaces. Here, $a(\lambda)$ is a scalar function of the complex spectral parameter $\lambda$. We now substitute this ansatz into the $(d,2k,m)$-gYBE,
\begin{equation}
R_{1\cdots 2k}(\lambda_1)~ R_{1+m\cdots 2k+m}(\lambda_2)~ R_{1\cdots 2k}(\lambda_3)= R_{1+m\cdots 2k+m}(\lambda_3) ~ R_{1\cdots 2k}(\lambda_2)~ R_{1+m\cdots 2k+m}(\lambda_1)\;,
\end{equation}
and expanding both sides, we obtain 
\begin{eqnarray}
    &\left[a(\lambda_1)+a(\lambda_3)\right]~\mathcal{M}_{1\cdots 2k}+a(\lambda_2)~\mathcal{M}_{1+m\cdots 2k+m}+ a(\lambda_1)a(\lambda_3)~\mathcal{M}_{1\cdots 2k}^2&\nonumber\\&+a(\lambda_1)a(\lambda_2)a(\lambda_3)~\mathcal{M}_{1\cdots 2k}\mathcal{M}_{1+m\cdots 2k+m}\mathcal{M}_{1\cdots 2k}& \nonumber\\[2mm]
    &=\left[a(\lambda_1)+a(\lambda_3)\right]~\mathcal{M}_{1+m\cdots 2k+m}+a(\lambda_2)~\mathcal{M}_{1\cdots 2k}+ a(\lambda_1)a(\lambda_3)~\mathcal{M}_{1+m\cdots 2k+m}^2&\nonumber\\&+a(\lambda_1)a(\lambda_2)a(\lambda_3)~\mathcal{M}_{1+m\cdots 2k+m}\mathcal{M}_{1\cdots 2k}\mathcal{M}_{1+m\cdots 2k+m}&
\end{eqnarray}
To obtain non-trivial solutions of the gYBE, we require the operator $\mathcal{M}_{1\cdots 2k}$ to anti-commute with its shifted neighbor $\mathcal{M}_{1+m\cdots 2k+m}$ and obeys the involution condition, $\mathcal{M}^2=\kappa\,\mathbb{1}$. These constraints lead to the following functional equation 
\begin{equation}
    a(\lambda_2) = \frac{a(\lambda_1)+ a(\lambda_3)}{1-\kappa\,a(\lambda_1)a(\lambda_3)}\;.
\end{equation}
This equation suggests that the $R$-matrix in \eqref{eq:R-gYBE} satisfies the additive form [$\lambda_2=\lambda_1+\lambda_3$] of the gYBE with $a(\lambda)$,
\begin{equation}
    a(\lambda) = \frac{1}{\sqrt{\kappa}}\tan{\lambda}\;,
\end{equation}
where $\alpha$ is an arbitrary non-zero constant.

\section{Derivation of the $(d,6,3)$-Majorana Hamiltonian}
\label{app:derivation-Hamiltonian}
In this appendix, we present the integrable derivation of the Hamiltonian and the corresponding translation operator from the $RTT$-relation \eqref{eq:RTT}. The key ingredient is the transfer matrix $\tau(\lambda)$, defined as \eqref{eq:transfermatrix}
\begin{eqnarray}
    \tau(\lambda)={\rm tr}_{\vec{\alpha},\vec{\beta}}\left[T_{\vec{\alpha}}(\lambda)\right]={\rm tr}_{\vec{\alpha},\vec{\beta}}\left[T_{\vec{\beta}}(\lambda)\right].
\end{eqnarray}
Here ${\rm tr}_{\vec{\alpha},\vec{\beta}}[\,\cdot\,]$ denotes performing partial trace over the auxiliary Hilbert space ${\cal H}_{\vec{\alpha},\vec{\beta}}$, which carries the representation  $\varrho_{\vec{\alpha},\vec{\beta}}$ of the auxiliary Majorana modes $\gamma_{\alpha_k},\gamma_{\beta_k},~k=1,2,3$. Further assume that the physical Majorana modes, $\gamma_j,~j=1,\cdots,6N$, have the representation $\varrho_{\rm P}$ on a physical Hilbert space ${\cal H}_{\rm P}$. For most of our discussion, however, we will adopt the notation $\Gamma_j:=\varrho_{\rm P}(\gamma_j)$. One can then construct the global representation of all the Majorana modes on the Hilbert space ${\cal H}:={\cal H}_{\vec{\alpha},\vec{\beta}}\otimes{\cal H}_{\rm P}$ as 
\begin{eqnarray}
    &\varrho(\gamma_{\alpha_k})=\varrho_{\vec{\alpha},\vec{\beta}}(\gamma_{\alpha_k})\otimes{\mathbb 1}_{\rm P},\qquad \varrho(\gamma_{\beta_k})=\varrho_{\vec{\alpha},\vec{\beta}}(\gamma_{\beta_k})\otimes{\mathbb 1}_{\rm P},\qquad k=1,2,3,\nonumber\\
    &\varrho(\gamma_j)={\rm i}\varrho_{\vec{\alpha},\vec{\beta}}(\gamma_{\alpha_1}\gamma_{\alpha_2}\gamma_{\alpha_3}\gamma_{\beta_1}\gamma_{\beta_2}\gamma_{\beta_3})\otimes \varrho_{\rm P}(\gamma_j),\qquad j=1,\cdots,6N.
\end{eqnarray}
Note that, the auxiliary Majorana modes act trivially on the physical Hilbert space, while the physical Majorana modes have non-trivial action on the auxiliary Hilbert space. This very property captures the non-local nature of the Majorana algebra. Now one can use the fact that different Majoranas anticommute with each other to show that
\begin{eqnarray}
    {\rm tr}_{\vec{\alpha},\vec{\beta}}[\varrho_{\vec{\alpha},\vec{\beta}}(\gamma_{\alpha_j})]={\rm tr}_{\vec{\alpha},\vec{\beta}}[\varrho_{\vec{\alpha},\vec{\beta}}(\gamma_{\alpha_j})\varrho_{\vec{\alpha},\vec{\beta}}(\gamma_{\alpha_k})]={\rm tr}_{\vec{\alpha},\vec{\beta}}[\varrho_{\vec{\alpha},\vec{\beta}}(\gamma_{\alpha_j})\varrho_{\vec{\alpha},\vec{\beta}}(\gamma_{\alpha_k})\varrho_{\vec{\alpha},\vec{\beta}}(\gamma_{\alpha_l})]=0,
\end{eqnarray}
for $j\neq k\neq l$. As a corollary, the $\tilde{R}_{\vec{\alpha},\vec{\beta}}(\lambda-\mu)$-operator in \eqref{eq:RTT} acts trivially on the physical Hilbert space. Hence one can multiply the $RTT$-relation \eqref{eq:RTT} by $\tilde{R}_{\vec{\alpha},\vec{\beta}}^{-1}(\lambda-\mu)$ from either left or right and can take the partial trace over ${\cal H}_{\vec{\alpha},\vec{\beta}}$ to obtain
\begin{eqnarray}
    [\tau(\lambda),\tau(\mu)]=0.
\end{eqnarray}
Since the monodromy operators $T_{\vec{\alpha}}(\lambda)$ and $T_{\vec{\beta}}(\lambda)$ act on the global Hilbert space ${\cal H}$, taking the partial trace of the above monodromy operators over the auxiliary Hilbert space ${\cal H}_{\vec{\alpha},\vec{\beta}}$ yields the transfer matrix $\tau(\lambda)$, which act only on the physical Hilbert space ${\cal H}_{\rm P}$. This completes the proof of the commutativity of the transfer matrices \eqref{eq:transfermatrix}.

It is instructive to begin with the translation operator, which can be found by considering the transfer matrix $\tau(\lambda)$ at $\lambda=0$ as 
\begin{eqnarray}
    \tau(0) &=& {\rm tr}_{\vec{\alpha},\vec{\beta}}\left[\tilde{R}_{\vec{\alpha},\vec{N}}(0)\cdots \tilde{R}_{\vec{\alpha},\vec{1}}(0)\right]={\rm tr}_{\vec{\alpha},\vec{\beta}}\left[\left(P^-_{\alpha_1,3N-2}P^-_{\alpha_2,3N-1}P^-_{\alpha_3,3N}\right) \cdots \left(P^-_{\alpha_1,1}P^-_{\alpha_2,2}P^-_{\alpha_3,3}\right)\right].
\end{eqnarray}
With some effort, one can show that
\begin{eqnarray}
    \tau(0)&\simeq& \Gamma_1\Gamma_2\Gamma_3\prod_{j=1}^{6N-3}\left(\frac{\Gamma_j-\Gamma_{j+3}}{\sqrt{2}}\right).
\end{eqnarray}
To obtain the Hamiltonian, we differentiate the transfer matrix $\tau(\lambda)={\rm tr}_{\vec{\alpha},\vec{\beta}}\left[T_{\vec{\alpha}}(\lambda)\right]$ with respect to the spectral parameter $\lambda$ and set $\lambda=0$ as
\begin{eqnarray}
   \frac{{\rm d}\tau(\lambda)}{{\rm d}\lambda}\Biggr|_{\lambda=0}&=& \sum\limits_{j=1}^{2N} {\rm tr}_{\vec{\alpha},\vec{\beta}}\left[\left(P^-_{\alpha_1,6N-2}P^-_{\alpha_2,6N-1}P^-_{\alpha_3,6N}\right)\cdots\left(P^-_{\alpha_1,3j-2}P^-_{\alpha_2,3j-1}P^-_{\alpha_3,3j}\right)\right.\nonumber\\ 
    &&\quad\left.(\gamma_{\alpha_1}+\gamma_{\alpha_2}+\gamma_{\alpha_3})\gamma_{3j-2}\gamma_{3j-1}\gamma_{3j}\left(P^-_{\alpha_1,3j-5}P^-_{\alpha_2,3j-4}P^-_{\alpha_3,3j-3}\right)\cdots\left(P^-_{\alpha_1,1}P^-_{\alpha_2,2}P^-_{\alpha_3,3}\right) \right]\nonumber\\
    &=&{\rm tr}_{\vec{\alpha},\vec{\beta}}\left[T_{\vec{\alpha}}(0)(\gamma_{\alpha_1}+\gamma_{\alpha_2}+\gamma_{\alpha_3})\gamma_{1}\gamma_{2}\gamma_{3} \right]+\tau(0)\sum_{j=2}^{2N}\left(\Gamma_{3j-5}+\Gamma_{3j-4}+\Gamma_{3j-3}\right)\Gamma_{3j-2}\Gamma_{3j-1}\gamma_{3j}. 
\end{eqnarray}
After some simple algebraic manipulations, we get
\begin{eqnarray}
    {\rm tr}_{\vec{\alpha},\vec{\beta}}\left[T_{\vec{\alpha}}(0)(\gamma_{\alpha_1}+\gamma_{\alpha_2}+\gamma_{\alpha_3})\gamma_{1}\gamma_{2}\gamma_{3} \right]=-\tau(0)(\Gamma_{6N-2}+\Gamma_{6N-1}+\Gamma_{6N})\Gamma_{1}\Gamma_{2}\Gamma_{3}.
\end{eqnarray}
We now can define the integrable Hamiltonian as 
\begin{eqnarray}
   H= \tau(0)^{-1}\frac{{\rm d}\tau(\lambda)}{{\rm d}\lambda}\bigg|_{\lambda=0}&=&\sum\limits_{j=1}^{2N}(-1)^{\delta_{j,1}}(\omega_1\Gamma_{3j-5}+\omega_2\Gamma_{3j-4}+\omega_3\Gamma_{3j-3})\Gamma_{3j-2}\Gamma_{3j-1}\Gamma_{3j}=\sum_{j=1}^{2N}(-1)^{\delta_{j,2N}}{\cal M}_j.
\end{eqnarray}
For this derivation to go through we require the total number of Majorana fermions to be an even number. Therefore, we work with $6N$ number of Majorana modes. Moreover, it is clear that the inclusion of the parameters $(\omega_1,\omega_2,\omega_3)$ does not alter the derivation, and the local Hamiltonian density is identified as $\mathcal{M}_j(\Omega)$. Note that, by a slight abuse of notation, we use $\gamma$ in the main text instead of $\Gamma$ introduced here in order to avoid introducing too many new symbols.

\section{Boost operator formalism}
\label{app:Boost-derivation}
In this appendix, we construct the boost operator for integrable models associated with the $(d,6,3)$-gYBE. We use the composite notation $\vec{j}=(3j-2,3j-1,3j)$ throughout the derivation. The derivative of the equation \eqref{eq:6-3-gYBE-non-braided} with respect to $\mu$ at $\mu=0$ is, 
\begin{eqnarray}\label{eq:d(gYBE)}   
\tilde{R}_{\vec{1},\vec{2}}(\lambda)\tilde{R}_{\vec{1},\vec{3}}'(\lambda)\tilde{R}_{\vec{2},\vec{3}}(0)+\tilde{R}_{\vec{1},\vec{2}}{(\lambda)}\tilde{R}_{\vec{1},\vec{3}}{(\lambda)}\tilde{R}_{\vec{2},\vec{3}}'{(0)}=\tilde{R}_{\vec{2},\vec{3}}'{(0)}\tilde{R}_{\vec{1},\vec{3}}{(\lambda)}\tilde{R}_{\vec{1},\vec{2}}{(\lambda)}+\tilde{R}_{\vec{2},\vec{3}}{(0)}\tilde{R}_{\vec{1},\vec{3}}'{(\lambda)}\tilde{R}_{\vec{1},\vec{2}}{(\lambda)}.\nonumber\\
\end{eqnarray}
From the expression of the $R$-matrix \eqref{eq:(6,3)-R-1+3}, we observe that
\begin{equation*}
    \tilde{R}_{\vec{1},\vec{2}}'{(0)}= \Pi_{\vec{1},\vec{2}}~ \mathcal{M}_{\vec{1}},\quad\tilde{R}_{\vec{1},\vec{2}}{(0)}=\Pi_{\vec{1},\vec{2}},\qquad \Pi_{\vec{1},\vec{2}}=P^-_{1,4}P^-_{2,5}P^-_{3,6}. 
\end{equation*}
Substituting these relations into \eqref{eq:d(gYBE)}, and simplifying, we obtain
\begin{eqnarray}\label{eq:preboost}
    \left[\mathcal{M}_{\vec{2}},\tilde{R}_{\vec{1},\vec{3}}(\lambda)\tilde{R}_{\vec{1},\vec{2}}(\lambda) \right]=\tilde{R}_{\vec{1},\vec{3}}(\lambda)\tilde{R}'_{\vec{1},\vec{2}}(\lambda)-\tilde{R}'_{\vec{1},\vec{3}}(\lambda)\tilde{R}_{\vec{1},\vec{2}}(\lambda).
\end{eqnarray}
At this stage, one must account for the passage of the Majorana permutation operator $\Pi$ through the $R$ matrices. However, this does not modify the final result. Since each $R$ matrix contains only Majorana-parity-even (quartic) terms within $\mathcal{M}$, no additional minus signs arise under permutation. Although commuting a $\Pi$ through another $\Pi$ inside $\tilde{R}$ may in principle generate a sign, the relevant expression involves crossing two such $R$ matrices, leading to an exact cancellation. One may therefore impose the standard exchange relations on the physical indices.
We now relabel the Majorana fermions according to 
\begin{eqnarray}
    \gamma_1\to \gamma_{\alpha_1},~2\to \gamma_{\alpha_2},~\gamma_{3}\to \gamma_{\alpha_3},\quad \gamma_4\to \gamma_{3j-2},~\gamma_5\to \gamma_{3j-1},~\gamma_6\to \gamma_{3j},\quad \gamma_7\to \gamma_{3j+1},~\gamma_8\to \gamma_{3j+2},~\gamma_9\to \gamma_{3j+3}
,
\end{eqnarray}
which can be written in a more compact way as
\begin{eqnarray}
    \vec{1}\to \vec{\alpha},\quad \vec{2}\to\vec{j},\quad \vec{3}\to\vec{j+1},\qquad \vec{\alpha}=(\alpha_1,\alpha_2,\alpha_3),\quad \vec{j}=(3j-2,3j-1,3j).
\end{eqnarray}
Here $\vec{\alpha}=(\alpha_1,\alpha_2,\alpha_3)$ denotes the auxiliary Majorana indices. Now we can multiply the resulting expression by the index $j$ and sum over $j$ from $-\infty$ to $\infty$, leading to
\begin{eqnarray}\label{eq:preboost2}
     &&\sum \limits_jj\left[\mathcal{M}_{{j}}\,,\,\tilde{R}_{\vec{\alpha},\vec{j+1}}\tilde{R}_{\vec{\alpha},\vec{j}} \right]=
     \sum\limits_j j\left(\tilde{R}_{\vec{\alpha},\vec{j+1}}\tilde{R}'_{\vec{\alpha},\vec{j}}-\tilde{R}'_{\vec{\alpha},\vec{j+1}}\tilde{R}_{\vec{\alpha},\vec{j}}\right).
\end{eqnarray}
Let us now multiply the above equation by the string $\prod_{m=\infty}^{j+2} \tilde{R}_{\vec{\alpha},\vec{m}}$ from left and by the string $\prod_{k=j-1}^{-\infty} \tilde{R}_{\vec{\alpha},\vec{k}}$ from right. Then the right hand side of the above equation simplifies to 
\begin{eqnarray}\label{eq:b-1}
    \sum\limits_j j\prod\limits_{m=\infty}^{j+1} \tilde{R}_{\vec{\alpha},\vec{m}}\left(\tilde{R}'_{\vec{\alpha},\vec{j}}\right)\prod\limits_{k=j-1}^{-\infty} \tilde{R}_{\vec{\alpha},\vec{k}}-\sum\limits_j j \prod\limits_{m=\infty}^{j+2} \tilde{R}_{\vec{\alpha},\vec{m}}\left(\tilde{R}'_{\vec{\alpha},\vec{j+1}}\right)\prod\limits_{k=j}^{-\infty} \tilde{R}_{\vec{\alpha},\vec{k}}.
\end{eqnarray}
We now relabel the index $j$ to $j+1$ in the first term of \eqref{eq:b-1} and obtain
\begin{eqnarray}
    \sum_j \prod\limits_{m=\infty}^{j+2} \tilde{R}_{\vec{a},\vec{m}}(\lambda)\left(\tilde{R}'_{\vec{a},\vec{j+1}}(\lambda)\right)\prod\limits_{k=j}^{-\infty} \tilde{R}_{\vec{a},\vec{k}}(\lambda)=\frac{{\rm d}T_{\vec{a}}(\lambda)}{{\rm d}\lambda},\qquad T_{\vec{\alpha}}(\lambda)=\prod_{j=-\infty}^\infty\tilde{R}_{\vec{\alpha},\vec{j}}(\lambda),
\end{eqnarray}
where the monodromy operator $T_{\vec{\alpha}}$ describes an infinitely extended fermionic chain. 
Combining the above expression with the left hand side of \eqref{eq:preboost2}, we finally arrive at
\begin{eqnarray}
    \frac{{\rm d}T_{\vec{\alpha}}(\lambda)}{{\rm d}\lambda}=\left[\sum \limits_j j\mathcal{M}_{j},T_{\vec{\alpha}}(\lambda)\right].
\end{eqnarray}
One now can take partial trace over the auxiliary Hilbert space, i.e., ${\rm tr}_{\vec{\alpha},\vec{\beta}}\left[T_{\vec{\alpha}}(\lambda)\right]=\tau(\lambda)$ and arrive at an exact well-known dynamical expression of the transfer matrix $\tau(\lambda)$ in terms of the \textit{boost operator} $\mathcal{B}$, namely  
\begin{equation}\label{eq:boost}
    \frac{{\rm d}\tau(\lambda)}{{\rm d}\lambda}= \left[\mathcal{B}, \tau(\lambda)\right], \qquad \mathcal{B}=\sum_{j=-\infty}^{\infty} j\mathcal{M}_{j},\qquad \tau(\lambda)=\tau(0)~ \exp\left[{\sum\limits_{k=1}^{\infty}\lambda^kI_{k+1}}\right].
\end{equation}
It is now possible to determine the higher order conserved charges from the relation \eqref{eq:boost}, which is presented by
\begin{eqnarray}
    I_{r+1}=\frac{1}{r}\left[\mathcal{B},I_r\right], ~~r>1.
\end{eqnarray}
The conserved quantities associated with our model are clearly written down in the main text.

\section{Multi-site Hamiltonians associated to $k=5$}
\label{app:k=5}
In each of the instances below, the $\mathcal{M}$'s are normalized to square to identity after dividing by an appropriate factor as explained in Section \ref{sec:gYBE-HFF}.
The 10, $\gamma$ Majoranas can be split into either 2, 4, 6, 8 or 10 parts. The only possible split when there are 2 parts is $1+1$. The  $\mathcal{M}$ operators are then given by 
\begin{eqnarray}
    \mathcal{M}_{1\cdots 10} = \begin{cases}
\left(\sum\limits_{j=1}^5~\omega_j\gamma_j\right)\left(\sum\limits_{j=6}^{10}~\omega_j\gamma_j\right), \\
\left(\omega_1\gamma_1\gamma_2\gamma_3+\omega_2\gamma_4 + \omega_3\gamma_5 \right)\left(\omega_1\gamma_6\gamma_7\gamma_8+\omega_2\gamma_9 + \omega_3\gamma_{10} \right),
    \end{cases}
\end{eqnarray}
with the $\omega_j's\in\mathbb{R}$. Note that Majorana fermion parity in each part of the split is preserved. For higher values of $k$, we expect to obtain more choices for possible parity preserving operators constructed out of the $\gamma$ Majoranas. This is is seen with the second choice of $\mathcal{M}$ in this case. We will see more such occurrences below. 

For 4 parts the splits can be either $1+3$ and its partner $3+1$ or $2+2$. We will write down the $\mathcal{M}$'s for just $1+3$ and $2+2$. For the former, these are,
\begin{eqnarray}
    \mathcal{M}_{1\cdots 10} =\begin{cases}
    \left(\sum\limits_{j=1}^5~\omega_j\gamma_j\right)\left(\omega_1\gamma_6 + \omega_2\gamma_7 + \omega_3\gamma_8\right)\gamma_9\gamma_{10}, \\
       \left(\sum\limits_{j=1}^5~\omega_j\gamma_j\right) \left(\omega_1\gamma_6 + \omega_2\gamma_7\right) \left(\omega_3\gamma_8 + \omega_4\gamma_9\right)\gamma_{10},\\
       \left(\omega_1\gamma_1\gamma_2\gamma_3 + \omega_2\gamma_5\right)\left( \omega_3\gamma_6 + \omega_4\gamma_7\right)\gamma_9\gamma_{10}.
    \end{cases}
\end{eqnarray}
And for the latter these are given by,
\begin{eqnarray}
    \mathcal{M}_{1\cdots 10} = \begin{cases}
\left(\sum\limits_{j=1}^4~\omega_j\gamma_j\right)\gamma_5\left(\sum\limits_{j=6}^9~\omega_j\gamma_j\right)\gamma_{10}~;~\sum\limits_{j=1}^4\omega_j\omega_{j+5}=0, \\
  \left(\omega_1\gamma_1\gamma_2\gamma_3 + \omega_2\gamma_4\right)\gamma_5\left(\omega_3\gamma_6\gamma_7\gamma_8 + \omega_4\gamma_9\right)\gamma_{10}~;~\omega_1\omega_3 + \omega_2\omega_4=0, \\
\left(\sum\limits_{j=1}^3~\omega_j\gamma_j\right)\left(\sum\limits_{j=4}^5~\omega_j\gamma_j\right)\left(\sum\limits_{j=6}^8~\omega_j\gamma_j\right)\left(\sum\limits_{j=9}^{10}~\omega_j\gamma_j\right).
    \end{cases}
\end{eqnarray}
For the last choice, the conditions on the coefficients $\omega$'s are 
\begin{eqnarray}
    &\omega_2\omega_5\omega_8\omega_9 + \omega_3\omega_4\omega_7\omega_{10} - \omega_2\omega_4\omega_8\omega_{10} - \omega_3\omega_5\omega_7\omega_9 = 0,\nonumber\\
    &\omega_1\omega_4\omega_6\omega_9 + \omega_1\omega_5\omega_6\omega_{10} + \omega_{10}\omega_4\omega_7\omega_9 + \omega_2\omega_5\omega_7\omega_{10} + \omega_3\omega_4\omega_8\omega_9 + \omega_3\omega_5\omega_8\omega_{10}=0, \nonumber \\
    &\omega_1\omega_5\omega_7\omega_6 + \omega_2\omega_4\omega_6\omega_{10} - \omega_1\omega_4\omega_7\omega_{10} - \omega_2\omega_5\omega_6\omega_9=0=
    \omega_1\omega_5\omega_8\omega_9 + \omega_3\omega_4\omega_6\omega_{10} - \omega_3\omega_5\omega_6\omega_{9} - \omega_1\omega_4\omega_8\omega_{10}.
\end{eqnarray}
We do not show the $\mathcal{M}$ operators corresponding to the $3+1$ split, as they are similar to the $1+3$ split.
Next we break $\mathcal{M}$ into 6 parts with possible splits being $1+5$ or $5+1$ and $3+3$. For the case of the $1+5$ split we have just two possible $\mathcal{M}$'s given by
\begin{eqnarray}
    \mathcal{M}_{1-10} = \begin{cases}
\left(\sum\limits_{j=1}^{5}\omega_j\gamma_j\right)\prod\limits_{j=6}^{10}~\gamma_j, \\
\left(\omega_1\gamma_1\gamma_2\gamma_3 + \omega_2\gamma_4 + \omega_3\gamma_5 \right)\prod\limits_{j=6}^{10}~\gamma_j.
    \end{cases}
\end{eqnarray}
When the split is $3+3$, the $\mathcal{M}$'s are given by
\begin{eqnarray}
  \mathcal{M}_{1\cdots 10} = \begin{cases}
\left(\sum\limits_{j=1}^3\omega_j\gamma_j\right)\gamma_4\gamma_5\left(\sum\limits_{j=1}^3\omega_j\gamma_{j+5}\right)\gamma_9\gamma_{10}, \\
\left(\omega_1\gamma_1+\omega_2\gamma_2\right)\left(\omega_3\gamma_3+\omega_4\gamma_4\right)\gamma_5\left(\omega_6\gamma_6+\omega_7\gamma_7\right)\left(\omega_8\gamma_8+\omega_9\gamma_9\right)\gamma_{10}.  
  \end{cases}  
\end{eqnarray}
For the latter we require additional constraints on the coefficients $\omega$ given by 
\begin{eqnarray}
    \omega_2\omega_4\omega_6\omega_9 + \omega_2\omega_3\omega_6\omega_8 - \omega_1\omega_3\omega_7\omega_8 - \omega_1\omega_4\omega_7\omega_9 =0=
    \omega_1\omega_4\omega_6\omega_8 + \omega_2\omega_4\omega_7\omega_8 - \omega_1\omega_3\omega_6\omega_9 - \omega_2\omega_3\omega_7\omega_9.
\end{eqnarray}
There are no possible $\mathcal{M}$ operators for the $4+2$ or $2+4$ splits.
The operator $\mathcal{M}$ can be decomposed into 8 parts with the splits, $5+3$ or $3+5$ and $4+4$. The former has two solutions for $\mathcal{M}$ given by,
\begin{eqnarray}
    \mathcal{M}_{1\cdots 10} = \begin{cases}
\prod\limits_{j=1}^5~\gamma_j\left(\sum\limits_{j=1}^3\omega_j\gamma_{j+5}\right)\gamma_9\gamma_{10}, \\
\prod\limits_{j=1}^5~\gamma_j\left(\omega_1\gamma_6+\omega_2\gamma_7 \right)\left(\omega_3\gamma_8+\omega_4\gamma_9\right)\gamma_{10}.
    \end{cases}
\end{eqnarray}
The only solution type for $\mathcal{M}$ of the $4+4$ split is,
\begin{eqnarray}
    \mathcal{M}_{1\cdots 10} = \left(\omega_1\gamma_1 + \omega_2\gamma_2 \right)\gamma_3\gamma_4\gamma_5\left(\omega_3\gamma_6 + \omega_4\gamma_7 \right)\gamma_8\gamma_9\gamma_{10},\quad\omega_1\omega_3+\omega_2\omega_4=0.
\end{eqnarray}
More solutions of this type can be obtained by shifting the factor in the $\left(\cdot\right)$. For instance $\mathcal{M}$ can also be realized as 
\begin{eqnarray}
    \mathcal{M}_{1\cdots 10} = \gamma_1\left(\omega_1\gamma_2 + \omega_2\gamma_3 \right)\gamma_4\gamma_5\gamma_6\left(\omega_3\gamma_7 + \omega_4\gamma_8 \right)\gamma_9\gamma_{10},\quad\omega_1\omega_3+\omega_2\omega_4=0,
\end{eqnarray}
and so on. 
Finally we have the $5+5$ split when $\mathcal{M}$ is broken into 10 parts. The only solution for this is given by
\begin{eqnarray}
    \mathcal{M}_{1\cdots 10}= \prod\limits_{j=1}^{10}~\gamma_j.
\end{eqnarray}
We now present an example of the Hamiltonian corresponding to the one of the above multi-site Majorana operator $\mathcal{M}$,
\begin{eqnarray}
    \mathcal{M}_{1-10}^{(2+2)}=\left(\omega_1\gamma_1\gamma_2\gamma_3 + \omega_2\gamma_4\right)\gamma_5\left(\omega_3\gamma_6\gamma_7\gamma_8 + \omega_4\gamma_9\right)\gamma_{10},\quad\omega_1\omega_3 + \omega_2\omega_4=0.
\end{eqnarray}
Following the same procedure outlined in the Appendix \ref{app:derivation-Hamiltonian} [differing only in the definition of the monodromy matrix, which is now supported on five auxiliary indices], we obtain the Hamiltonian corresponding to the above $\mathcal{M}$ operator as
\begin{eqnarray}
    H^{(2+2)}_{k=5}=\sum\limits_{j} \left(\omega_1\gamma_{5j-4}\gamma_{5j-3}\gamma_{5j-2} + \omega_2\gamma_{5j-1}\right)\gamma_{5j}\left(\omega_3\gamma_{5j+1}\gamma_{5j+2}\gamma_{5j+3} + \omega_4\gamma_{5j+4}\right)\gamma_{5j+5}.
\end{eqnarray}
The spin-1/2 representation of this Hamiltonian takes the form 
\begin{eqnarray}
    H^{(2+2)}_{k=5}&=&\sum\limits_{j}(\omega_1 Z_{5j-4}Y_{5j-3}+\mathrm{i}\omega_2 X_{5j-3} )(-\omega_3 Y_{5j}+\mathrm{i}\omega_4 X_{5j-2}Z_{5j})\nonumber\\&&+(-\omega_1 X_{5j-2}Y_{5j}+\mathrm{i}\omega_2 Z_{5j} )(\omega_3 Z_{5j+1}Y_{5j+2}+\mathrm{i}\omega_4 X_{5j+2})X_{5j+3}.
\end{eqnarray}
It is clear that this expression describes a periodic spin-chain Hamiltonian satisfying the periodic condition $5N+j \equiv j$.

\end{document}